\documentclass[10pt,journal,compsoc]{IEEEtran}
\usepackage{makecell}
\usepackage{hyperref}
\usepackage{array}
\usepackage{graphicx,amssymb,amsmath}
\usepackage{enumitem}
\usepackage{amsfonts}
\usepackage{subcaption}
\usepackage{multicol}
\usepackage[noadjust]{cite}
\usepackage{setspace}
\usepackage{graphicx}
\usepackage{float}
\usepackage {url}
\usepackage{stfloats}
\usepackage{amsthm,pifont}
\usepackage{flushend}
\usepackage{cases,subeqnarray}
\usepackage{bm,multirow,bigstrut}
\usepackage{amsmath, amsthm, amssymb}
\usepackage{textcomp}
\usepackage{latexsym,bm}
\usepackage{booktabs}
\usepackage{xcolor}
\usepackage{mathtools}
\usepackage{dsfont}
\usepackage{extarrows}
\usepackage{epsfig}
\usepackage{epsfig}
\usepackage{epstopdf}
\usepackage{algpseudocode}
\usepackage{algorithmicx,algorithm}
\usepackage{svg}
\theoremstyle{plain}

\theoremstyle{plain}

\usepackage{amsmath}

\IEEEoverridecommandlockouts
\usepackage{fancyhdr}

\begin{document}
\title{{Reinforcement Learning With LLMs Interaction For Distributed Diffusion Model Services}}
\author{Hongyang~Du$^*$, Ruichen Zhang, Dusit~Niyato,~\IEEEmembership{Fellow,~IEEE}, Jiawen~Kang, Zehui Xiong, Shuguang~Cui,~\IEEEmembership{Fellow,~IEEE}, Xuemin~(Sherman)~Shen,~\IEEEmembership{Fellow,~IEEE}, and Dong~In~Kim,~\IEEEmembership{Fellow,~IEEE}
\thanks{H. Du is with the Department of Electrical and Electronic Engineering, University of Hong Kong, Pok Fu Lam, Hong Kong SAR, China (e-mail: duhy@eee.hku.hk).}
\thanks{R. Zhang, and D. Niyato are with the School of Computer Science and Engineering, Nanyang Technological University, Singapore (e-mail: ruichen.zhang@ntu.edu.sg, dniyato@ntu.edu.sg).}
\thanks{J. Kang is with the School of Automation, Guangdong University of Technology, China. (e-mail: kavinkang@gdut.edu.cn)}
\thanks{Z. Xiong is with the Pillar of Information Systems Technology and Design, Singapore University of Technology and Design, Singapore (e-mail: zehui\_xiong@sutd.edu.sg)}
\thanks{S. Cui is with the School of Science and Engineering (SSE) and the Future Network of Intelligence Institute (FNii), Chinese University of Hong Kong (Shenzhen), China (e-mail: shuguangcui@cuhk.edu.cn).}
\thanks{X.~Shen is with the Department of Electrical and Computer Engineering, University of Waterloo, Waterloo, ON N2L 3G1, Canada (e-mail: sshen@uwaterloo.ca).}
\thanks{D.~I.~Kim is with the Department of Electrical and Computer Engineering, Sungkyunkwan University, Suwon 16419, South Korea (email:dongin@skku.edu).}
\thanks{Corresponding authors: Hongyang Du.}
}
\fancypagestyle{firstpage}{
	\fancyhf{}  
	\lhead{%
		\begin{minipage}{0.9\textwidth}
			\includegraphics[height=0.9cm]{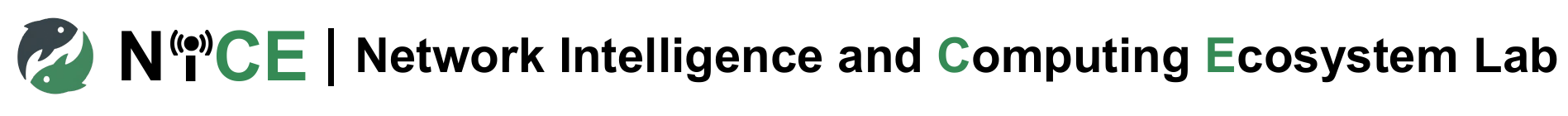}%
		\end{minipage}
	}
	\renewcommand{\headrulewidth}{0pt}
}
\maketitle
\thispagestyle{firstpage}
\maketitle

\begin{abstract}
Distributed Artificial Intelligence-Generated Content (AIGC) has attracted {\color{black}significant attention, but two key challenges remain: maximizing subjective Quality of Experience (QoE) and improving energy efficiency, which are particularly pronounced in widely adopted Generative Diffusion Model (GDM)-based image generation services.} In this paper, we propose a novel user-centric Interactive AI (IAI) approach for service management, with a distributed GDM-based AIGC framework {\color{black}that emphasizes efficient and cooperative deployment. The proposed method restructures the GDM inference process by allowing users with semantically similar prompts to share parts of the denoising chain.} Furthermore, to maximize the users' subjective QoE, we propose a IAI approach, i.e., Reinforcement Learning With Large Language Models Interaction (RLLI), which utilizes Large Language Model (LLM)-empowered generative agents to replicate users interaction, providing real-time and subjective QoE {\color{black}feedback aligned with diverse user personalities.} Lastly, we present the GDM-based Deep Deterministic Policy Gradient (G-DDPG) algorithm, adapted to the proposed RLLI framework, {\color{black}to allocate communication and computing resources effectively while accounting for subjective user traits and dynamic wireless conditions. Simulation results demonstrate that G-DDPG improves total QoE by $15\%$ compared with the standard DDPG algorithm.}
\end{abstract}
\IEEEpeerreviewmaketitle

\section{Introduction}
{\color{black}\IEEEPARstart{T}{he} growing demand for Artificial Intelligence-Generated Content (AIGC) services in multimedia and business applications~\cite{guo2022systematic} is driven by the advancement of Generative AI (GenAI) models, which provide scalable and consistent outputs in both text and imagery~\cite{du2023enabling}. ChatGPT, for example, reached over 100 million active users within two months~\cite{lund2023chatting}, demonstrating its influence on text-based interactions.}
In the realm of visual content generation, Stable Diffusion's capacity to create images from text prompts shows significant progress in multi-modal technologies~\cite{stabdiff,croitoru2023diffusion}. 
The widespread adoption of AIGC services in human societies indicates a notable transition to Interactive AI (IAI) as the next evolutionary phase of GenAI~\cite{amershi2019guidelines}. {\color{black}This shift is reshaping human-content interaction and reflects ongoing developments in human-AI collaboration.}

\begin{figure}[t]
\centering
\includegraphics[width=0.48\textwidth]{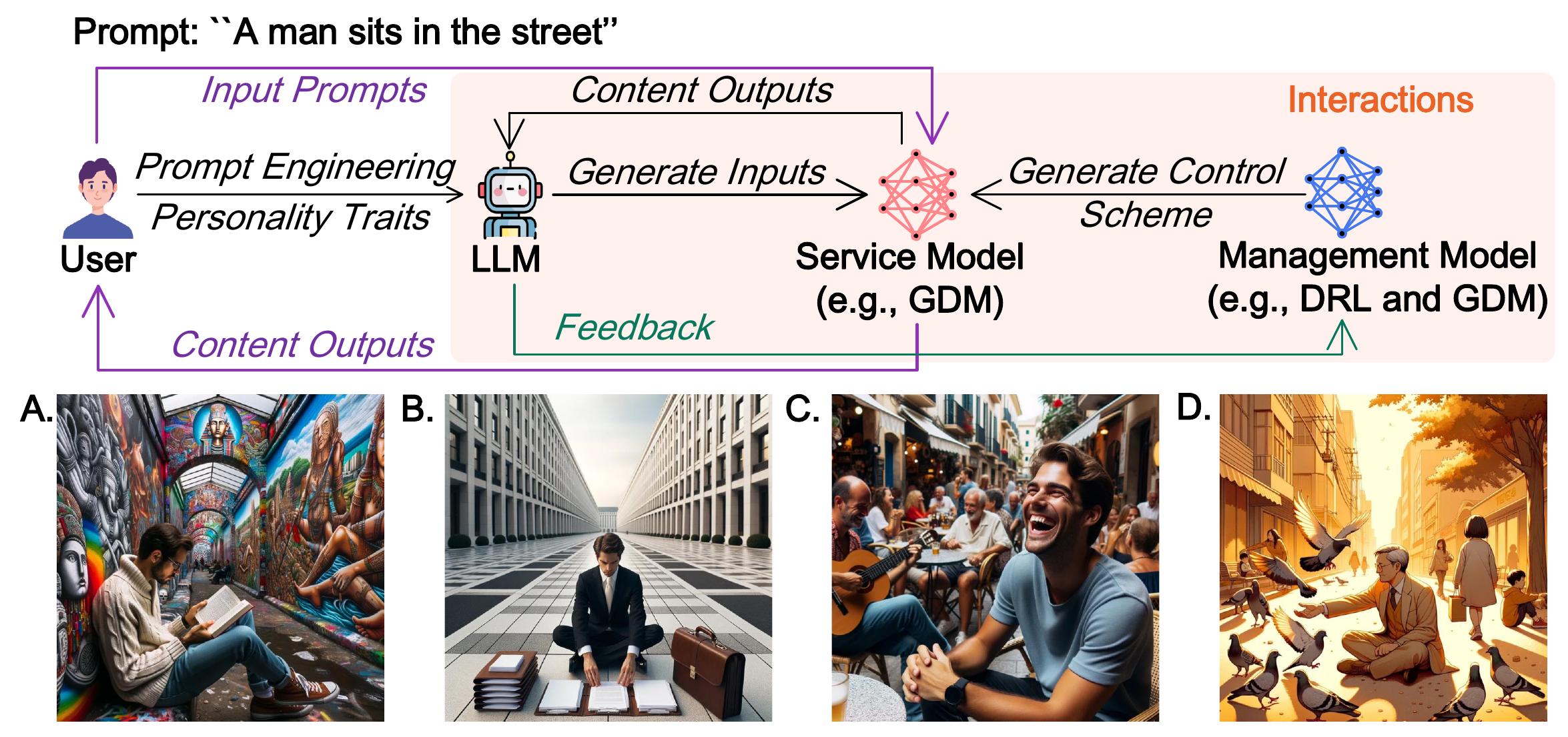}%
\caption{The basic framework of IAI and four images generated with the prompt {\textit{``A man sits in the street''}}. {\textbf{Part A}} is a man engrossed in a book against vibrant street art appeals to users with high {\textit{openness}}. {\textbf{Part B}} is a formally dressed man on a clean street, resonating with users high in {\textit{conscientiousness}}. {\textbf{Part C}} is a sociable street cafe scene, suitable for users with high {\textit{extraversion}}. {\textbf{Part D}} is a man feeding pigeons in a peaceful setting with playing children, fitting for users with high {\textit{agreeableness}}.}
\label{qoec}
\end{figure}
\begin{figure*}[t]
\centering
\includegraphics[width=0.98\textwidth]{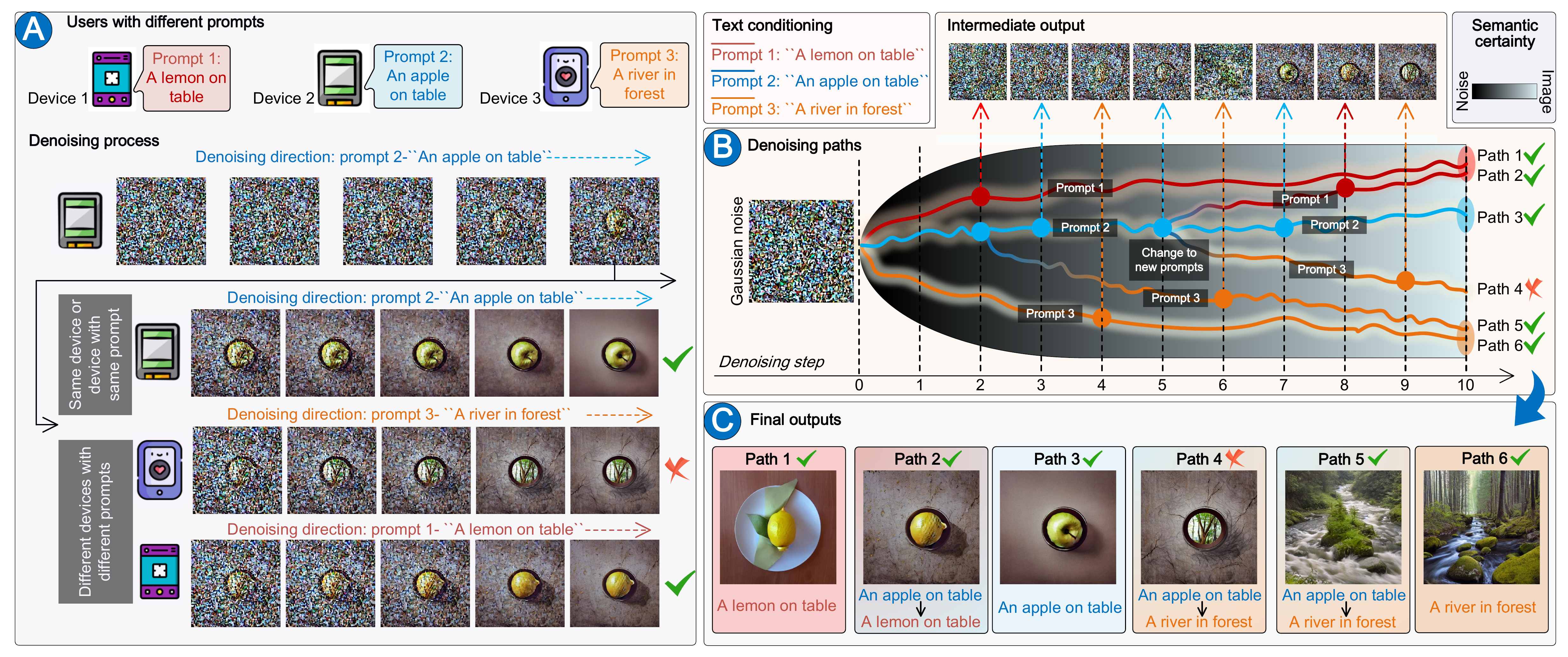}%
\caption{The working principle of the GDM and motivations behind distributed denoising inference process. {\textbf{Part A}} depicts the cooperative inference process across devices where, starting with Gaussian noise on {\textit{Device 2}}, it denoises using {\textit{Prompt 2}} before {\textit{Devices 1}} and $3$ continue in succession towards their respective prompts. {\textbf{Part B}} shows the fundamentals of distributed GDM inference, illustrating how denoising path directions alter with changing prompts, and emphasizing the high semantic similarity of {\textit{Prompt 1}} to {\textit{Prompt 2}}, contrasting it with the distance between {\textit{Prompt 2}} and {\textit{Prompt 3}}. Consequently, Path 2 aligns with {\textit{Prompt 1}}'s requirements, unlike {\textit{Path 4}} with{\textit{ Prompt 3}}. {\textbf{Part C}} showcases the final generation outcomes for all five paths.}
\label{diffthree}
\end{figure*}

However, Quality of Experience (QoE) maximization in AIGC services emerges as a critical challenge, due to the subjective nature of human perception, which extends beyond objective image quality metrics~\cite{picard2001toward,juluri2015measurement}. 
Four images in Fig.~\ref{qoec} exemplify this complexity, showing four distinct images that are generated under a same prompt {\textit{``A man sits in the street''}}. {\color{black}Although each image is of high quality, they align differently with user personality profiles, resulting in varying QoE evaluations. Therefore, network designers aim to develop service and management models that adapt to individual personalities to improve QoE. However, the lack of a precise mathematical representation of QoE makes this optimization challenging.}
While some studies have adopted psychological laws to approximate users' subjective QoE~\cite{du2023attention}, these methods often {\color{black}fail to capture the complexity of real-world scenarios. An alternative is to train management models using Reinforcement Learning with Human Feedback (RLHF), which relies on expert-provided QoE labels. This method is expensive, ethically sensitive, and difficult to implement in real-time, raising the first research question:}

{\textit{{\textbf{C1.}} How to obtain human-aware subjective QoE feedback efficiently and design the communication and computing resource allocation algorithm?}}

Addressing {\textit{{\textbf{C1}}} requires the exploration of cutting-edge approaches, among which IAI stands out as a promising solution~\cite{amershi2019guidelines}. 
IAI focuses on designing AI models that learn and adapt through user interaction, progressively advancing AI models' performance and operational efficacy. 
This paradigm shift, from static to dynamic learning systems, equips IAI-enhanced networks with the ability to offer tailored responses and proactively adapt, marking a significant stride in personalized AI-based service managements.
For AIGC services, as shown in Fig.~\ref{qoec}, we propose Reinforcement Learning With LLM Interaction (RLLI) algorithm as one step towards IAI, incorporates Deep Reinforcement Learning (DRL) for its suitability in dynamic environments~\cite{luong2019applications} and Large Language Model (LLM) for advanced knowledge understanding and generative capabilities:
\begin{itemize}
\item DRL regards user QoE as quantifiable rewards and circumvents complexity of mathematically modeling subjective QoE. When integrated with GenAI techniques such as Generative Diffusion Model (GDM)~\cite{du2023beyond} that effectively handle complex data modeling and produce high-quality samples, DRL can be further enhanced to allocate network resources efficiently while factoring in user satisfaction.
\item LLM-empowered Ggenerative Agents (GAs) can represent real AIGC users with various personalities to generate QoE feedback, minimizing human resource input and associated ethical risks. By embedding human personality traits into GAs through prompts~\cite{liu2023pre}, these GAs can simulate human interaction in the training of DRL algorithms.
\end{itemize}

{\color{black}Furthermore, AIGC models impose substantial energy demands during the inference phase~\cite{croitoru2023diffusion}.
Particularly in GDM-based image generation, multiple denoising steps are essential for producing high-quality outputs. This iterative process progressively removes noise, resulting in considerable energy and time consumption~\cite{croitoru2023diffusion}.
Although few-step models have been proposed to reduce computation and accelerate inference~\cite{luo2023latent,sauer2023adversarial}, they often degrade image quality and detail compared to multi-step approaches~\cite{geng2023visual,du2023demofusion}. In practice, widely used AIGC systems such as Stable Diffusion~\cite{stabdiff} and Sora~\cite{sora} continue to depend on multi-step denoising, underscoring its necessity for achieving high-quality results. Additionally, transmitting these high-quality images, particularly in latency-sensitive applications like augmented reality and telemedicine, requires significant network bandwidth~\cite{xu2023unleashing}, placing increased demands on future Sixth-Generation (6G) networks that aim to support human-centric and high QoE services~\cite{dang2020should}. 
These challenges lead to the second research question:}

{\textit{{\textbf{C2.}} How to harness network capabilities to facilitate energy-efficient and low-latency GDM inference?}}

Fortunately, advancements in model compression facilitate the deployment of large GenAI models directly on network edge devices~\cite{ma2023llm,kim2023architectural}. 
This forms a foundation for network-assisted AIGC, particularly collaborative distributed diffusion model~\cite{du2023exploring}, to address {\textit{{\textbf{C1}}}.
As shown in Part A of Fig.~\ref{diffthree}, by distributing the denoising steps across multiple network edge devices, we can achieve a more flexible and efficient network resource utilization.
More importantly, generation tasks with semantically similar prompts can share several denoising steps, thereby saving overall energy and time consumption. 
As shown in Part B of Fig.~\ref{diffthree}, the denoising {\textit{Path 2}}, which starts with ``{\textit{Prompt 2: An apple on table}}'' and switches to ``{\textit{Prompt 1: A lemon on table}}'' can still converge nearly with {\textit{Path 1}} that uses ``{\textit{Prompt 1}}'' throughout.
We can observe from Part C of Fig.~\ref{diffthree} that when employing an appropriate distributed denoising process, such as switching prompts in an early stage or maintaining semantic similarity between prompts, high-quality images can be generated that is corresponding to user prompts while reducing overall resource consumption. 
For example, we can observe a clear reduction in energy and time costs when generating images via {\textit{Path 2}} and {\textit{Path 3}}, compared to independent operations through {\textit{Path 1}} and {\textit{Path 3}}.

{\color{black}Note that the distributed GDM-based AIGC framework enables semantic fusion of prompts from multiple users, making the quality of generated images dependent on the designed communication and computing resource allocation scheme. This dependency is addressed by the proposed edge-based IAI solution, which introduces a new optimization dimension to enhance AIGC service quality and improve user QoE.} The contributions of this paper are summarized as follows.
\begin{itemize}
\item We introduce a collaborative distributed GDM-based AIGC framework for image generation services. This approach enhances user privacy by generating the final content on edge devices, reducing the total network energy and time overheads (for {\textit{{\textbf{C1}}}}).
\item We present a QoE feedback scheme by using LLM-empowered GAs to simulate the human's different personalities. With the aid of prompts and assigning one agent per user, GAs can mimic users of diverse subjective preferences, delivering evaluations of quality of generated images (for {\textit{{\textbf{C2}}}}).
\item We propose an IAI algorithm, i.e., RLLI, with LLM-generated QoE at its core. The goal is to determine the optimal denoising steps, i.e., computing energy consumption, at server and user devices, and downlink transmission power, i.e., communications energy consumption, of the server to user devices, based on the user personalities and wireless conditions (for {\textit{{\textbf{C2}}}}).
\end{itemize}
The rest of this paper is organized as follows: Section~\ref{S2} reviews related literature. In Section~\ref{S3}, we describes the system model and formulates the problem for our distributed GDM-based AIGC framework. Section~\ref{S4} details the RLLI algorithm.  Section~\ref{S5} conducts a evaluation of the proposed framework and RLLI algorithm. The paper concludes with Section~\ref{S6}. A list of mathematical symbols and functions most frequently used in this paper is available in Table~\ref{table1xxx}.

\begin{table}[t]
	\caption{Mathematical Notations}
	\vspace{-0.2cm}
	\centering
	\label{table1xxx}
	{\small
		\begin{tabular}{m{0.7cm}|m{7.3cm}}
			\toprule
			${\bm{c}}_k$ & The prompt from the $k$-${\rm{th}}$ user, where $ k = 1,\ldots,K $ \\
			$\mathcal{D}$ & The decoder that decodes a latent representation in the latent space into an image in the RGB space \\
			$\delta_0$ & The energy cost for each denoising step of the server \\
			$\delta_k$ & The energy cost for each denoising step of the $k$-${\rm{th}}$ device \\
			$E_{\rm T}$ & The energy budget in the server \\
			$E_{{\rm{T}}_k}$ & The energy budget in the $k$-${\rm{th}}$ device \\
			$\mathcal{E}$ & The encoder that encodes an image in the RGB space into a latent representation in the latent space \\
			${{\bm{g}}_k}$ & The wireless channel gains from the server to the $k$-${\rm{th}}$ device \\
			$K$ & The total number of users \\
			$P_k$ & The transmission power from the server to the $k$-${\rm{th}}$ device \\
			$\mathcal{Q}_k$ & The QoE of the $k$-${\rm{th}}$ user \\
			$\mathcal{Q}_{\rm th}$ & The QoE threshold \\
			$t_0$ & The number of denoising steps in the server \\
			$t_k$ & The number of denoising steps in the $k$-${\rm{th}}$ device \\
			$\tau\left( \cdot \right)$ & The domain-specific encoder that projects ${\bm{c}}$ to a latent representation \\
			${\bm{u}_k}$ & The personality of the $k$-${\rm{th}}$ user \\
			${{\bm{\varepsilon}}_\theta }$ & The denoising network of the GDM \\
			\bottomrule
	\end{tabular}}
\end{table}

\begin{table}[t]
\caption{{\color{black} List of Abbreviations}}
\vspace{-0.2cm}
\centering
\label{table:abbreviations}
{\color{black} {\small
\begin{tabular}{m{1.8cm}|m{6.2cm}}
\toprule
\textbf{Abbreviation} & \textbf{Full Term} \\
\midrule
AIGC & Artificial Intelligence-Generated Content \\
QoE & Quality of Experience \\
GDM & Generative Diffusion Model \\
IAI & Interactive AI \\
VIT & Visual Instruction Tuning \\
LLM & Large Language Model \\
RLLI & Reinforcement Learning With LLM Interaction \\
DRL & Deep Reinforcement Learning \\
RLHF & Reinforcement Learning with Human Feedback \\
DDPG & Deep Deterministic Policy Gradient \\
G-DDPG & GDM-based Deep Deterministic Policy Gradient \\
GA & Generative Agent \\
BEP & Bit Error Probability \\
SNR & Signal-to-Noise Ratio \\
\bottomrule
\end{tabular}}}
\end{table}

\section{Related Work}~\label{S2}
In this section, we discuss several related works, including GDM, aesthetics analysis, and LLM.
\subsection{Generative Diffusion Models}
{\color{black}GDMs are essential to various AIGC services. Their applications include image and video generation in computer vision using DDPM~\cite{ho2020denoising} and video diffusion models~\cite{ho2022video}, text generation and editing through Diffusion-LM~\cite{li2022diffusion} and DiffusER~\cite{reid2023diffuser}, audio synthesis with ProDiff~\cite{huang2022prodiff} and DiffWave~\cite{kongdiffwave}, as well as specialized tasks such as graph generation~\cite{niu2020permutation} and molecular structure creation~\cite{ketata2023diffdock}. As privacy-sensitive applications grow, there is increasing interest in deploying GDMs on edge devices. While model compression has made this feasible, the inference process—particularly the iterative denoising steps—remains energy-intensive, increasing the total energy cost across the network. However, the sequential structure of the denoising chain enables distributed computation. Cooperative distributed GDM~\cite{du2023exploring} exploits this structure by allowing devices to share partial denoising steps, which reduces redundant computation and lowers network-wide energy consumption.}

\subsection{Aesthetics Analysis and Big Five Personality Traits}
{\color{black}Traditional approaches to Image Aesthetic Assessment (IAA) have focused on universal aesthetic scores, overlooking the subjective nature of aesthetic preferences driven by individual personality differences.}
Recent developments, such as those integrating the Big Five personality traits into a multi-task deep learning model~\cite{li2020personality} for IAA, represent a paradigm shift.
The Big Five personality traits~\cite{saucier1998beyond}, comprising openness, conscientiousness, extraversion, agreeableness, and neuroticism, {\color{black}provide a widely accepted structure for characterizing personality}. Specifically, openness captures creativity and a willingness of users to explore new experiences; conscientiousness involves diligence and reliability; extraversion denotes sociability and assertiveness; agreeableness reflects cooperativeness and kindness; and neuroticism is associated with emotional instability and anxiety.
{\color{black}Embedding these traits into IAA enables a transition from general scoring to individualized assessments, offering the potential to align AIGC models with users' psychological profiles and make AIGC services more personalized}~\cite{cristani2013unveiling}.

\subsection{LLM-empowered Generative Agents as Evaluators} 
{\color{black}LLMs-empowered GAs can interpret and respond to complex natural language instructions, making them a versatile interface for a range of tasks, including evaluative ones~\cite{wang2023chatgpt,lee2023rlaif}.}
Specifically, the authors in~\cite{wang2023chatgpt} explore the capabilities of ChatGPT in evaluating textual content across various human-aware criteria such as quality, tone, and coherence. 
These evaluations provide a foundation for extending LLM functionality into other domains. 
Subsequently, the authors in~\cite{ziems2023can} evaluate the potential of LLMs like ChatGPT in computational social science, examining their ability to perform both classification and generative tasks in a zero-shot manner. It is shown that LLMs demonstrate fair agreement with humans and can augment the annotation process.
More recently, the authors in~\cite{wang2023does} show that modern role-playing LLMs can successfully mimic specific personality traits, achieving an $82.8\%$ alignment with human perceptions.
{\color{black}These findings support the use of LLM-empowered GAs for simulating users with diverse subjective preferences in image evaluation tasks, positioning GA feedback as a viable alternative to human feedback for training service management algorithms.}

\section{System Model and Problem Formulation}\label{S3}
In this section, we analyze the principles and conditions under which GDMs can be executed separately on different devices. Furthermore, we discuss the deployment method and formulate the user-centric QoE optimization problem in our proposed distributed GDM-based AIGC framework.

\subsection{Distributed Generative Diffusion Model-based AIGC}
\begin{figure}[t]
\centering
\includegraphics[width=0.48\textwidth]{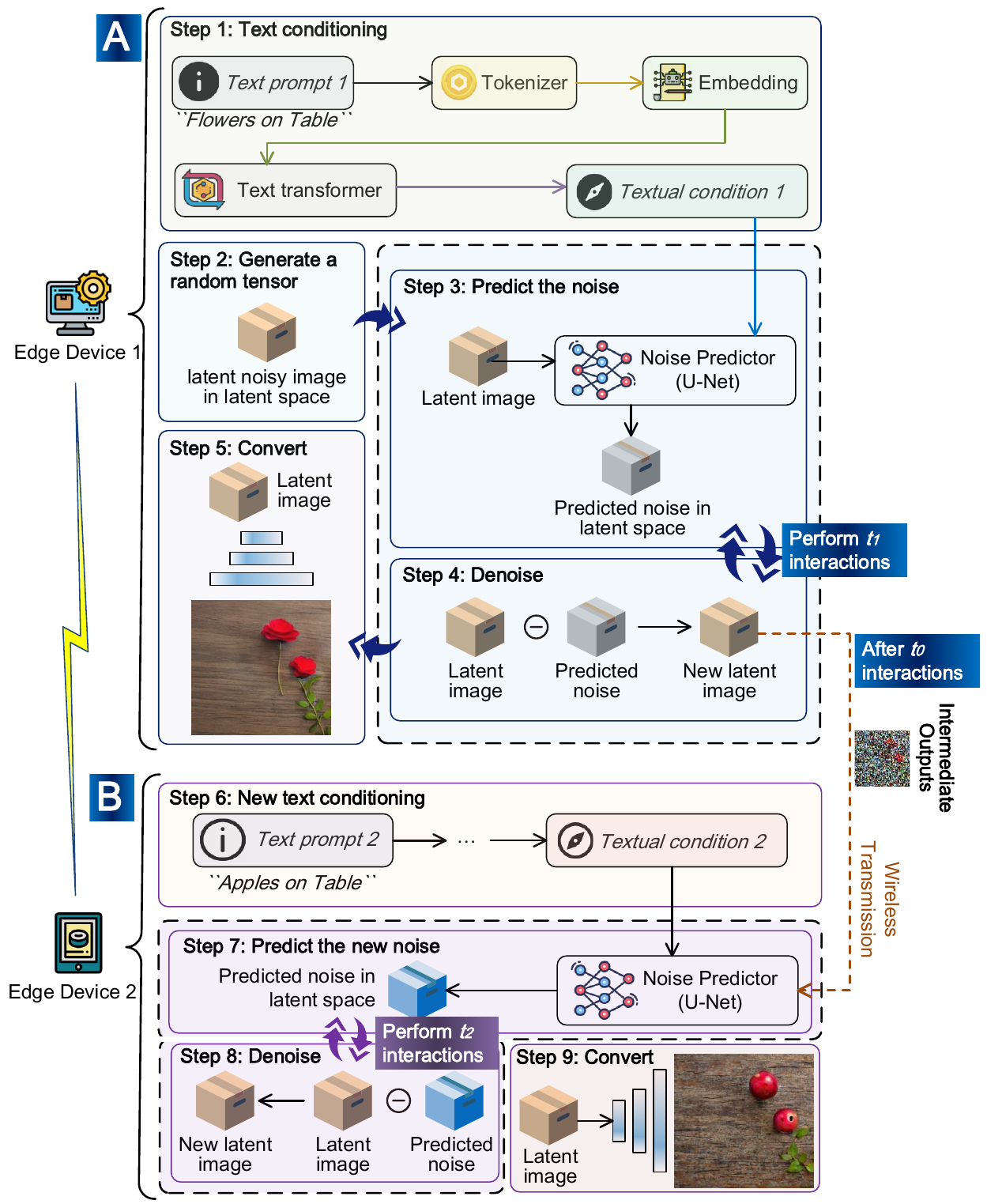}%
\caption{The working steps of GDM-based text-to-image generation service, and the principle to achieve multi-device distributed denoising process.}
\label{difftheory}
\end{figure}

\subsubsection{Latent Diffusion Models}
{\color{black}GDMs, as illustrated in Part A of Fig.~\ref{difftheory}, are probabilistic models that approximate a data distribution by iteratively denoising samples drawn from a normal distribution, effectively reversing a Markov chain over $T$ steps. The Stable Diffusion~\cite{stabdiff} stands out by adopting latent GDM, operating in the latent space rather than the high-dimensional image pixel space.}
Instead of manifesting a perturbed image, latent GDM introduces ``latent noise'', i.e., a stochastic tensor in the latent space, during training to perturb the latent representation. For example, an image with $512 \times 512$ resolution, the corresponding latent size is a much smaller, i.e., $4 \times 64 \times 64$, enabling faster processing and less memory usage.

Diving deeper into the mechanics, we consider ${\bm{x}}_0 \in {R^{H \times W \times 3}}$ is an image in Red, Green \& Blue (RGB) space. The encoder $\mathcal{E}$ encodes ${\bm{x}}_0$ into a latent representation as
\begin{equation}
{\bm{z}_0} = \mathcal{E}\left( {\bm{x}_0}\right),
\end{equation}
where ${\bm{z}_0} \in {R^{h \times w \times r}}$. The encoder $\mathcal{E}$ downsamples the image ${\bm{x}}_0$ by a factor $f$, i.e., $f = H/h = W/w$.
The decoder $\mathcal{D}$ reconstructs the image from the latent representation as
\begin{equation}
{{\tilde{\bm{x}}}_0} = \mathcal{D}\left({\bm{z}}_0\right)= \mathcal{D}\left(\mathcal{E}\left( {\bm{x}}_0\right)\right).
\end{equation}
GDMs can be specified in terms of a Signal-to-Noise Ratio (SNR) as
\begin{equation}
\gamma \left( t \right) = \frac{{\alpha _t^2}}{{\sigma _t^2}},
\end{equation}
where $\alpha _t$ and $\sigma _t$ are strictly positive scalar-value functions of $t$, and $\gamma \left( t \right)$ is strictly monotonically decreasing in time, i.e., $\gamma \left( t \right) < \gamma \left( s \right)$ for any $t > s$. 
We consider that both $\alpha _t$ and $\sigma _t$ are smooth, such that their derivatives with respect to $t$ are finite~\cite{sohl2015deep}. This diffusion process specification includes the variance-preserving diffusion process as a special case~\cite{sohl2015deep,ho2020denoising}, where ${\alpha _t}{ = }\sqrt {1 - \sigma _t^2} $. Another special case is the variance-exploding diffusion process~\cite{song2019generative,song2020score}, where ${\alpha _t}^2 = 1$. We use the variance-preserving version in this paper.
By starting from a data sample ${\bm{z}}_0$, a forward diffusion process $q$ can be defined as
\begin{equation}
q\left( {\left. {{{\bm{z}}_t}} \right|{{\bm{z}}_0}} \right) = {\cal N}\left( {\left. {{{\bm{z}}_t}} \right| {{\alpha _t}{{\bm{z}}_0}},\sigma _t^2{\bm{I}}} \right).
\end{equation}
The Markov structure for $t > s$ can be expressed as
\begin{equation}
q\left( {\left. {{{\bm{z}}_t}} \right|{{\bm{z}}_s}} \right) = {\cal N}\left( {\left. {{{\bm{z}}_t}} \right| {{\alpha _{\left. t \right|s}}{{\bm{z}}_0}},\sigma _{\left. t \right|s}^2{\bm{I}}} \right),
\end{equation}
where $ {\alpha _{\left. t \right|s}} = {\alpha _t}/{\alpha _s} $ and $ \sigma _{\left. t \right|s}^2 = \sigma _t^2 - \alpha _{\left. t \right|s}^2\sigma _s^2 $.

Denoising GDMs are generative models $p$ which revert the forward diffusion process with a similar Markov structure running backward as
\begin{equation}
p\left( {{{\bm{z}}_0}} \right) = p\left( {{{\bm{z}}_T}} \right)\prod\limits_{t = 1}^T {p\left( {\left. {{{\bm{z}}_{t - 1}}} \right|{{\bm{z}}_t}} \right)}.
\end{equation}
The prior $ p\left( {{{\bm{z}}_T}} \right) $ is typically choosen as a standard normal distribution. We parameterize ${p\left( {\left. {{{\bm{z}}_{t - 1}}} \right|{{\bm{z}}_t}} \right)}$ to specify it in terms of the true posterior ${q\left( {\left. {{{\bm{z}}_{t - 1}}} \right|{{\bm{z}}_t},{{\bm{z}}_0}} \right)}$. We use the network with parameter $\theta$, i.e., ${{\bm{\varepsilon}}}_\theta\left( {{{{\bm{z}}}_t},t} \right)$, to estimate the unknown original data sample ${{\bm{z}}}_0$. Then, we obtain that~\cite{kingma2021variational}
\begin{align}\label{de1}
p\left( {\left. {{{\bm{z}}_{t - 1}}} \right|{{\bm{z}}_t}} \right) & = q\left( {\left. {{{\bm{z}}_{t - 1}}} \right|{{\bm{z}}_t},{{\bm{z}}_0}} \right) \notag\\& 
= q\left( {\left. {{{\bm{z}}_{t - 1}}} \right|{{\bm{z}}_t},{{\bm{z}}_\theta }\left( {{{\bm{z}}_t},t} \right)} \right)
\notag\\&
= {\cal N}\left( {\left. {{{\bm{z}}_{t - 1}}} \right|{\mu _\theta }\left( {{{\bm{z}}_t},t} \right),\sigma _{\left. t \right|t - 1}^2\frac{{\sigma _{t - 1}^2}}{{\sigma _t^2}}{\bm{I}}} \right),
\end{align}
where 
\begin{equation}\label{de2}
{\mu _\theta }\left( {{{\bm{z}}_t},t} \right) = \frac{{{\alpha _{t\mid t - 1}}\sigma _{t - 1}^2}}{{\sigma _t^2}}{{\bm{z}}_t} + \frac{{{\alpha _{t - 1}}\sigma _{t\mid t - 1}^2}}{{\sigma _t^2}}{{\bm{\varepsilon}}_\theta }\left( {{{\bm{z}}_t},t} \right).
\end{equation}
We use the reparameterization \cite{ho2020denoising}
\begin{equation}\label{de3}
{{\bm{\varepsilon}} _\theta }\left( {{{\bm{z}}_t},t} \right) = \frac{{{{\bm{z}}_t} - {\alpha _t}{{\bm{z}}_\theta }\left( {{{\bm{\varepsilon}}_t},t} \right)}}{{{\sigma _t}}},
\end{equation}
to express the reconstruction term as a denoising objective. 

For convenience, we denote the denoising process that involves \eqref{de1}, \eqref{de2}, and \eqref{de3} as $\textit{Denoise}\left(\cdot\right) $ 
\begin{equation}
{\bm{z}}_{t-1} = \textit{Denoise}\left( {\bm{z}}_{t}, t\right).
\end{equation}
The corresponding loss function to train the latent GDM can be simplified to
\begin{equation}
{{\cal L}_{GDM}} = {E_{{\bm{z}},\varepsilon  \sim {\cal N}\left( {0,1} \right),t}}\left[ {\left\| {{{\varepsilon}}  - {{\bm{\varepsilon}} _\theta }\left( {{{\bm z}_t},t} \right)} \right\|_2^2} \right],
\end{equation}
with $t$ uniformly sampled from $ \left\{ {1, \ldots ,T} \right\} $. The neural backbone is realized as a time-conditional UNet~\cite{ronneberger2015u}. Since the forward process is fixed, $\bm{z}_t$ can be efficiently obtained from $\mathcal{E}$ during training, and samples from $p\left( z\right) $ can be decoded to image space with a single pass through $\mathcal{D}$.

\subsubsection{Conditioning Mechanisms}
Similar to other types of generative models~\cite{gauthier2014conditional,sohn2015learning}, GDMs are capable of modeling conditional distributions. 
This can be implemented with a conditional denoising auto-encoder ${{\bm{\varepsilon}}_\theta }\left( {{{\bm z}_t},t,{\bm{c}}} \right)$ to control the synthesis process through inputs ${\bm{c}}$ such as text~\cite{reed2016generative}, semantic maps~\cite{park2019semantic} or other image-to-image translation tasks~\cite{isola2017image}. 
To pre-process ${\bm{c}}$ from various modalities, such as language prompts, a domain-specific encoder is introduced that projects ${\bm{c}}$ to a latent representation $\tau\left({\bm{c}}\right)$. 
Based on image-conditioning pairs, we then train the conditional latent GDM via
\begin{equation}
{{\cal L}_{LDM}} = {E_{{\bm{z}},\varepsilon \sim {\cal N}\left( {0,1} \right),t, {\bm{c}}}}\left[ {\left\| {{{\varepsilon}}  - {{\bm{\varepsilon}} _\theta }\left( {{{\bm z}_t},t,\tau\left({\bm{c}}\right)} \right)} \right\|_2^2} \right].
\end{equation}
In the inference process, the denoising functions can be updated accordingly, by adding ${\bm{c}}$ as the condition together with step information $t$ as
\begin{equation}\label{falje}
{\bm{z}}_{t-1} = \textit{Denoise}\left( {\bm{z}}_{t}, t, \tau\left({\bm{c}}\right)\right).
\end{equation}

\subsubsection{Distributed Diffusion Model Inference}\label{agfg4eag}
\begin{algorithm}[t]
\caption{Distributed latent GDM Inference Process Among $K$ Devices} 
\label{Algorithm_SDiffusion}
\hspace*{0.02in} {\bf Input:}
the textual prompt $ \left\{ {{{\bm{c}}_1}, \ldots ,{{\bm{c}}_K}} \right\} $ from devices $1, \ldots, K$, respectively  \\
\hspace*{0.02in} {\bf Output:}
Generated image ${\tilde{\bm{x}}}_0$ for any device $k$ $\left( {k = 2, \ldots ,K} \right) $
\begin{algorithmic}[1]
\State {\textit{\#\# In Device 1}}
\State Sample initial Gaussian noise ${\bm{z}}_T$ in the latent space
\For{$t = T$ to $t_{\rm off} + 1$}
\State ${\bm{z}}_{t-1} \leftarrow \text{Denoise}({\bm{z}}_{t}, t, \tau\left( \bm{c}_i\right) )$, i.e., \eqref{falje}
\EndFor
\State \textit{\#\# Wireless Transmission}
\State Transmit the ${\bm{z}}_{t_{\rm off}}$ from device 1 to devices $2, \ldots, K$
\State {\textit{\#\# In Device $k$ $\left( {k = 2, \ldots ,K} \right) $}}
\For{$t = t_{\rm off}$ to $1$}
\State ${\bm{z}}_{t-1} \leftarrow \text{Denoise}({\bm{z}}_{t}, t, \tau\left( \bm{c}_k\right) )$, i.e., \eqref{falje}
\EndFor
\State {\textit{\#\# Image Reconstruction}}
\State ${\tilde{\bm{x}}}_0 \leftarrow \mathcal{D}({\bm{z}}_0)$
\State \Return ${\tilde{\bm{x}}}_0$
\end{algorithmic}
\end{algorithm}

The distributed GDM inference process relies on sequential denoising guided by textual prompts. We analyze mathematically the conditions for changing textual prompts {\color{black}while still ensuring that the denoising results remain reliable.}
Specifically, we consider the latent space, in which the diffusion process takes place, to be the Euclidean space denoted as \( \mathcal{L} \). Every point in \( \mathcal{L} \) corresponds to a latent vector. The analysis includes the following steps:
\begin{enumerate}[leftmargin=*]
\item \textit{Defining Denoising Path and Semantic Distance}: For each textual prompt $\bm{c}$, the denoising path, evolving from the initial Gaussian noise to the targeted semantic representation, is represented by $P_{\bm{c}}: \left[T, 0\right] \rightarrow \mathcal{L}$. Here, $P_{\bm{c}}\left(T\right)$ represents the Gaussian noise point, i.e., the start point in Part B of Fig.~\ref{diffthree}, while $P_{\bm{c}}\left(0\right)$ corresponds to the latent representation of the final generated image under the prompt $\bm{c}$, i.e., the end point in Part B of Fig.~\ref{diffthree}.

The semantic distance between two prompts, $\bm{c}_i$ and $\bm{c}_j$, in the latent space is defined by the Euclidean distance between their respective denoising path endpoints as
\begin{equation}
d\left( \bm{c}_i, 0, \bm{c}_j, 0\right)  = \lVert P_{\bm{c}_i}\left(0\right) - P_{\bm{c}_j}\left(0\right) \rVert.
\end{equation}
The semantic distance provides a direct measure: if the semantic information of $\bm{c}_i$ and $\bm{c}_j$ are closely related, their corresponding semantic representations, i.e., latent vectors, and the denoising paths under their guidance, should also be proximate in $\mathcal{L}$. 
Consequently, a smaller $d\left( \bm{c}_i, 0, \bm{c}_j, 0\right)$ indicates that substituting $\bm{c}_i$ for $\bm{c}_j$ is less likely to degrade the quality of generated images.

\item {\textit{Semantic Analysis:}} Another key observation is that extending the denoising path, i.e., $P_{\bm{c}}\left(t\right)$ with $t$ from $T$ to $0$, progressively incorporates the semantic information of the textual prompt $\bm{c}$ into the path.
To measure the difference in semantic information encoded within the latent vectors at step $t$ between prompts $\bm{c}_i$ and $\bm{c}_j$, we employ the Kullback-Leibler (KL) divergence:
\begin{equation}\label{kl}
{\mathcal{K}}\left(p\left( \bm{z}_t|\bm{c}_i\right)  || p\left( \bm{z}_t|\bm{c}_j\right) \right) = \sum_{z \in \bm{z}_t} p\left( z|\bm{c}_i\right) \log \frac{p\left( z|\bm{c}_i\right) }{p\left( z|\bm{c}_j\right) }.
\end{equation}
At the beginning of the denoising process, i.e., $t = T$, the noise component dominates, leading to a latent space with high entropy due to the absence of any meaningful semantic information. Thus, both prompts would produce almost similar distributions for the latent vectors. 
As a result, the KL divergence, ${\mathcal{K}}$, would be small, suggesting minimal difference between the paths $P_{\bm{c}_i}$ and $P_{\bm{c}_j}$, which means that $d\left( \bm{c}_i, T, \bm{c}_j, T\right)$ is small for any $\bm{c}_i$ and $\bm{c}_j$.
With the progression of denoising, the influence of the prompts begins to introduce semantic information, leading to a reduction in entropy.

\item \textit{Switching Denoising Path}:
Assuming a denoising path have been progressed $t_{\rm{off}}$ steps under $\bm{c}_i$ to an intermediate point $P_{\bm{c}_i}(t_{\rm{off}})$, where $0 < t_{\rm{off}} < T$. Now, the flowing denoising path should cover the semantic distance:
\begin{equation}\label{dis1}
d\left( \bm{c}_i, t_{\rm{off}}, \bm{c}_i, 0\right) = \lVert P_{\bm{c}_i}(t_{\rm{off}}) - P_{\bm{c}_i}(0) \rVert.
\end{equation}
However, if we intend to switch the prompt from $\bm{c}_i$ to $\bm{c}_j$, the distance from $P_{\bm{c}_i}(t_{\rm{off}})$ to the end-point of $\bm{c}_j$ is:
\begin{equation}\label{dis2}
d\left( \bm{c}_i, t_{\rm{off}}, \bm{c}_j, 0\right) = \lVert P_{\bm{c}_i}(t_{\rm{off}}) - P_{\bm{c}_j}(0) \rVert.
\end{equation}
This distance $d\left( \bm{c}_i, t_{\rm{off}}, \bm{c}_j, 0\right)$ is crucial as it informs the extent of path correction needed to align the denoising process towards the semantics of $\bm{c}_j$. 
From the above analysis, we conclude the following conditions for our distributed GDM-based AIGC framework:
\begin{itemize}
\item The new prompt $\bm{c}_j$ should have close semantic information related to $\bm{c}_i$, reducing the difference between~\eqref{dis1} and~\eqref{dis2}.
\item The transition to $\bm{c}_j$ should be made in a relatively early stage of the denoising process, i.e., when ${\mathcal{K}}$ as given in~\eqref{kl} is small, to reduce the influence of $\bm{c}_i$ and allowing $\bm{c}_j$ to predominantly lead the denoising direction.
\end{itemize}
\item \textit{Image Reconstruction}:
The image \( \tilde{\bm{x}}_0 \) reconstructed from the latent representation $P_{c_j}\left( 0\right) $ reflects the semantic influence of $\bm{c}_j$:
\begin{equation}
\tilde{\bm{x}}_0 \approx \mathcal{D}\left( P_{{\bm{c}}_j}\left( 0\right) \right).
\end{equation}
The decoder $\mathcal{D}$ maps the latent representation back to the image space. A high quality in $\tilde{\bm{x}}_0$ means the generated image is a precise representation, effectively capturing the semantics associated with $\bm{c}_j$.
\end{enumerate}
The overall algorithm for distributed latent GDM inference process among $K$ devices is given as {\textbf{Algorithm}}~\ref{Algorithm_SDiffusion}.
Note that our framework efficiently supports both synchronous and asynchronous operations. In synchronous operations, intermediate denoised results are directly transmitted to edge users without needing storage. In asynchronous operations, we can use a caching mechanism to store denoised results, as discussed in Section~\ref{S6}.

\subsection{Deployment Method}
\begin{figure}[t]
	\centering
	\includegraphics[width=0.4\textwidth]{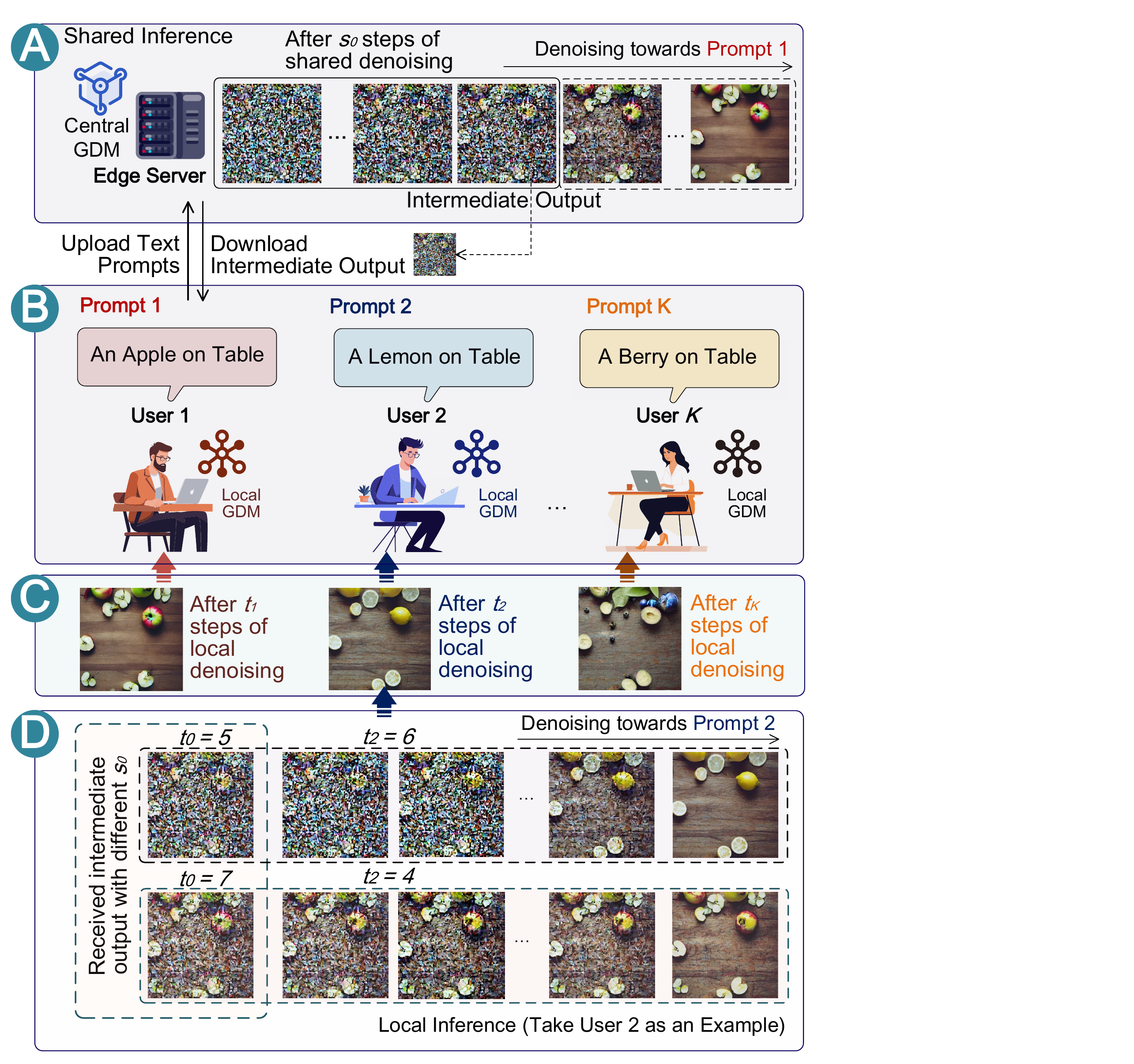}
	\vspace{-0.1cm}
	\caption{Deployment method for distributed GDM-based text-to-image AIGC services: {\textbf{Part A}} illustrates the shared inference mechanism where users with semantically similar prompts collaborate with the server. {\textbf{Part B}} highlights the diverse text-to-image requirements of each of the $K$ concurrent users. Following initial server processing, {\textbf{Part D}} delves into the distributed denoising process by each individual user. {\textbf{Part C}} presents the final generated images corresponding to user-specific prompts.}
	\label{model}
\end{figure}

{\color{black}The proposed distributed GDM-based AIGC framework leverages the flexible structure of the denoising process. By sharing intermediate denoised results across network devices, it reduces the system's overall energy consumption. Three network architectures are considered for deployment:}
\begin{enumerate}
	\item The \textit{Edge-to-Multiple Devices} architecture executes shared denoising steps at an edge server, and then transmitting intermediate denoised results to the user devices, as shown in Fig.~\ref{model}.
	\item The \textit{Device-to-Device} approach fosters a partnership between two devices, jointly executing denoising tasks, optimizing energy usage, and safeguarding user privacy, as shown in Fig.~\ref{difftheory}.
	\item The strategy of {\textit{Forming a Cluster among Multiple Devices}}, either with the assistance of an edge server or through self-organization, bolsters system adaptability and scalability to local requirements and environments, as shown in Part A of Fig.~\ref{diffthree}.
\end{enumerate}
{\color{black}These architectures introduce additional design considerations under real-world conditions, such as network variability and device heterogeneity. Transmission delays, errors, and device constraints, e.g., memory, should be accounted for to ensure stable performance. Our current analysis partially reflects these through channel modeling and energy cost formulations, but further system-level adaptations could be integrated as extensions. Without loss of generality, we focus on the \textit{Edge-to-Multiple Devices} architecture for the distributed GDM-based AIGC framework.}

Here, we consider a server executes $t_0$ denoising steps, followed by $K$ users, {\color{black}each executing} $t_k$ steps. Each {\color{black}server-side denoising step} incurs an energy cost $\delta_0$, while every user device has its unit energy cost $\delta_k$. Additionally, transmitting the intermediate result to the $k$-th user requires transmission power $P_k$, with a unit energy cost $\beta$.
{\color{black}Transmitting the intermediate result to the $k$-${\rm th}$ user requires transmission power $P_k$ with unit energy cost $\beta$. The QoE depends on both the allocation of communication and computing resources and the user's personality $\bm{u}_k$, which reflects aesthetic preferences and leads to subjective QoE evaluations. Here, the resource allocation variables for communication and computing are}
\begin{equation}\label{dseafa1}
	{\bm{t}} = \{t_0,t_1,\ldots,t_K\},
\end{equation}
and
\begin{equation}\label{dseafa2}
	{\bm{P}} = \{P_1,\ldots,P_K\}.
\end{equation}

The resource allocation variables, ${\bm{t}}$ and ${\bm{P}}$ , are significant in determining the efficacy of the AIGC service:
\begin{itemize}
	\item The transmission power, ${\bm{P}}$, impacts the bit error probability (BEP) during wireless transmission. Lower power levels risk degrading the integrity of the transmitted intermediate denoised results, potentially diminishing the quality of generated images.
	\item The number of denoising steps, ${\bm{t}}$, directly affects the quality of generated images~\cite{croitoru2023diffusion}. Specifically, the shared steps, $t_0$, set the initial latent representation. We dynamically and optimally adjust $t_0$ to balance efficiency and personalization, enhancing QoE while managing energy consumption.
	\item Notably, when transmission power is insufficient, which increases the bit errors in intermediate denoised results, adding more denoising steps can improve the final image quality~\cite{du2023exploring}. These additional steps function as an error-correction phase, reducing inaccuracies and improving the final result.
\end{itemize}

\subsection{User-centric QoE Maximization Problem}
The central challenge is maximizing subjective QoE across all users while considering user personalities, objective wireless conditions, energy constraints, and QoE thresholds. {\color{black}This integrated modeling approach captures two essential aspects of distributed AIGC services. First, using QoE as a utility function enables personalized content generation that aligns with user expectations. Second, incorporating energy constraints is necessary for efficient energy management across devices and network nodes. Mathematically, the objective and constraints are expressed as:}
\begin{align}\label{optimization}
	\mathop {\max }\limits_{\left\{ {{\bm{t}},{\bm{P}}} \right\}}  \quad & \sum_{k=1}^{K} \mathcal{Q}_k(t_0, t_k, P_k, \bm{u}_k), \\
	\textit{s.t.,} \quad 
	& \delta_k t_k \leq E_{{\rm{T}}_k}, \quad \forall k \in \{1, \dots, K\} \\
	& \mathcal{Q}_k(t_0, t_k, P_k, \bm{u}_k) \geq \mathcal{Q}_{\rm th}, \quad \forall k \in \{1, \dots, K\} \\
	& \delta_0 t_0 + \beta \sum_{k=1}^{K} P_k \leq E_{\rm{T}}.
\end{align}
{\color{black}The first constraint reflects the computing energy limitations of edge devices in executing denoising steps. The second constraint ensures that the AIGC service satisfies users' minimum QoE requirements. The third constraint, which is the most critical, addresses the overall energy budget and requires a trade-off between communication and computation. Specifically, increasing transmission power reduces the BEP, thereby improving the accuracy of transmitted intermediate denoised results. However, excessive transmission power reduces the energy available for performing denoising steps, which negatively impacts the final image quality. Conversely, if the transmission power is insufficient, even enhanced denoising at the edge device cannot fully correct the errors introduced during transmission.}

In the following, we use the RLLI method for the QoE maximization problem in our proposed distributed GDM-based AIGC framework.

\section{Reinforcement Learning With LLMs Feedback}\label{S4}
{\color{black}In this section, we present the RLLI method for evaluating the QoE of the distributed GDM framework described in Section~\ref{S3}. We first introduce an aesthetic-aware QoE model, then explain how LLM-empowered GAs provide subjective QoE feedback, which serves as the reward signal in DRL algorithms.}

\subsection{Aesthetic-aware QoE Modeling}
\begin{figure*}[t]
	\centering
	\includegraphics[width=0.8\textwidth]{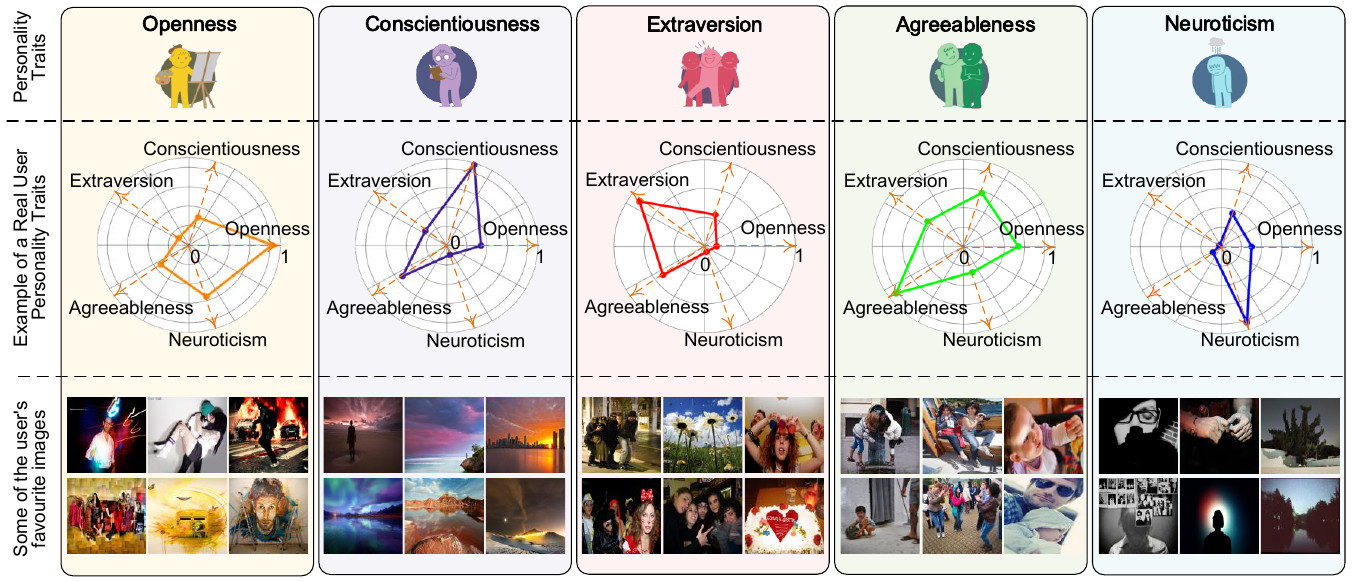}%
	\caption{Five types of personality traits in the Big-Five personality model: Openness, Conscientiousness, Extraversion, Agreeableness, and Neuroticism. We show real user personality scores from the PsychoFlickr database, along with examples of the images that they prefer.}
	\label{bigfive}
\end{figure*}
\subsubsection{Big Five Personality Traits}
{\color{black}Incorporating GAs into the DRL training process enables the integration of human subjective factors, particularly aesthetic preferences, which are critical to QoE in image-related AIGC services. Prior studies indicate that aesthetic preferences are influenced by individual personality traits~\cite{li2020personality,cristani2013unveiling,saucier1998beyond}. The Big Five personality traits~\cite{saucier1998beyond}, also known as the five-factor model, provide a widely accepted framework for characterizing personality differences}\footnote{One example of open source Big Five personality traits tests: \url{https://bigfive-test.com}.}. The Big Five model consists of five broad dimensions of personality traits including 
\begin{itemize}
\item {\textit{Openness:}} A trait characterized by a deep affinity for imagination, novel experiences, and a broad spectrum of interests.
\item {\textit{Conscientiousness:}} Denotes meticulousness and structure, often manifesting in methodical, goal-oriented behaviors.
\item {\textit{Extraversion:}} Embodies individuals who thrive in social interactions, drawing energy from a company of others.
\item {\textit{Agreeableness:}} Reflects proclivities towards trustworthiness, altruism, and prosocial behaviors.
\item {\textit{Neuroticism:}} Typified by emotional fluctuations and heightened sensitivity to environmental stressors and adversities.
\end{itemize}
{\color{black}Each of these traits represents a continuous spectrum of personality characteristics. Considering that recent studies have shown LLMs can effectively simulate the Big Five personality traits with an $82.8\%$ alignment to human perception~\cite{wang2023does}, we adopt this model to guide LLM-empowered GAs in mimicking AIGC service users' personalities~\cite{liu2023pre}. However, while personality traits provide a structured foundation for modeling user preferences, actual behavior may still vary in real-world contexts. We do not explicitly model such dynamics in this paper, but this represents a valuable direction for future refinement.}

In Fig.~\ref{bigfive}, we display five exemplary user profiles from the PsychoFlickr dataset~\cite{cristani2013unveiling}, {\color{black}showing their Big Five personality scores and selected preferred images. These profiles offer empirical evidence of the correlation between individual personality traits and aesthetic preferences in image selection.
This connection becomes especially important in the context of shared denoising within distributed GDMs. Unlike independent denoising, shared denoising allows semantically similar prompts to reuse intermediate diffusion steps, which improves efficiency but introduces interdependencies among outputs. As discussed in Section~\ref{agfg4eag}, the more steps that are shared, the more likely it is that the final images will influence one another. From a QoE perspective, this creates a potential conflict: users with distinct aesthetic preferences—shaped by their personalities—may experience degraded satisfaction if shared denoising alters the semantic or stylistic fidelity of their expected outputs.}
For instance, as depicted in Part B of Fig.~\ref{showllm}, one user, scoring high in agreeableness, might prefer an image generated from the prompt {\textit{``A dog sits in front of a bush''}} that has a well-groomed dog. 
However, if this prompt shares several shared denoising steps with another prompt such as {\textit{A tiger sits in front of a bush''}}, the final generated images for this user may have a dog that exhibits stylistic features of a tiger, causing the low QoE. 
Thus, the settings for shared denoising steps are pivotal in achieving maximum sum QoE.

\subsubsection{Prompt Design for LLM-empowered Generative Agents}
\begin{figure*}[t]
\centering
\includegraphics[width=0.8\textwidth]{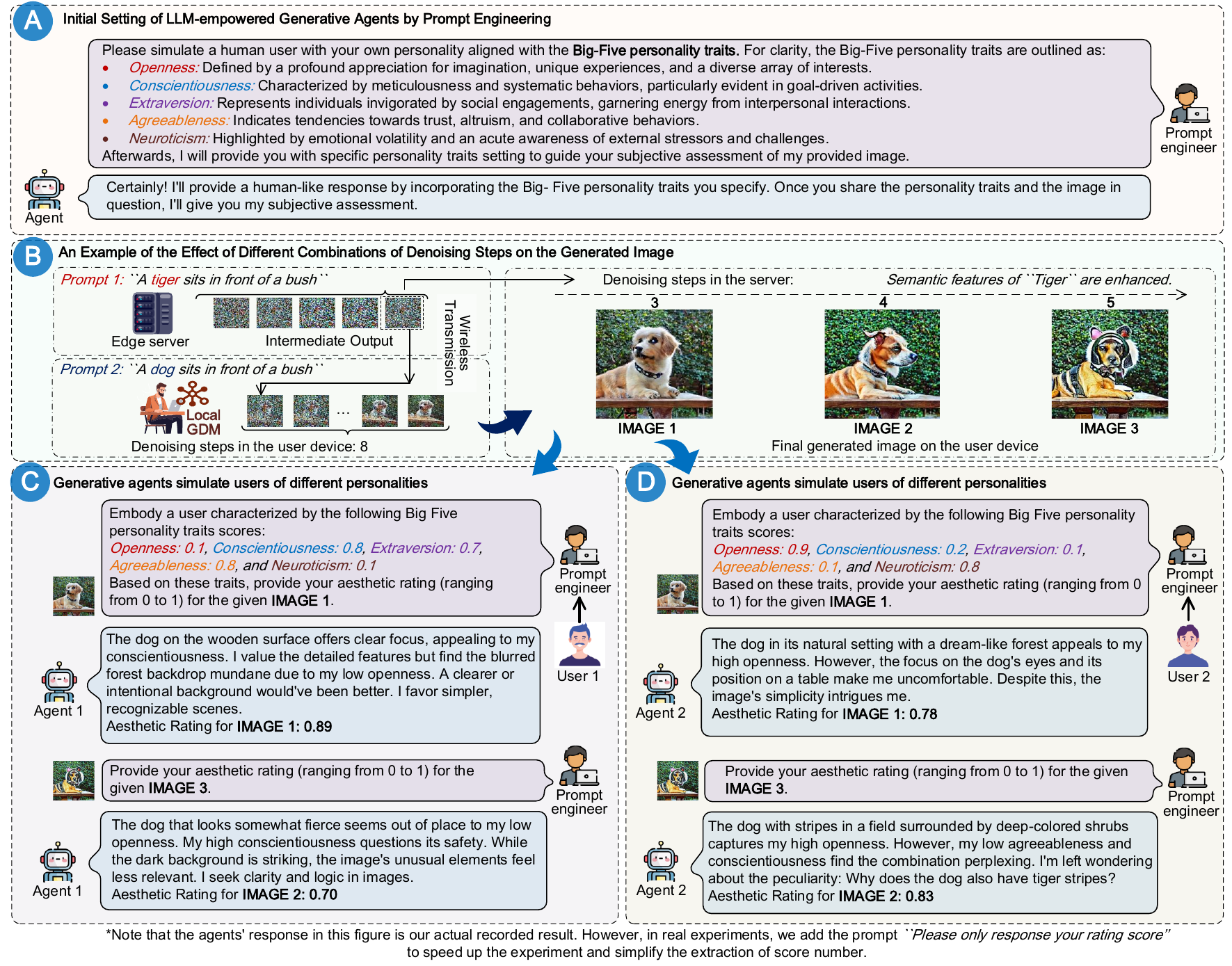}%
\caption{Prompts for LLM-empowered GAs settings. {\textbf{Part A}} illustrates the initial setup to acquaint the GAs with the Big-Five personality traits, preparing them for the subjective assessment task. {\textbf{Part B}} demonstrates a case where the different numbers of shared denoising steps lead to stylistic differences in the final generated images. {\textbf{Part C}} and {\textbf{Part D}} present two user personality trait configurations and the corresponding evaluated scores of generated images.}
\label{showllm}
\end{figure*}

The LLM-empowered GAs present a powerful mechanism to feedback human-aware subjective QoE values for the generated content.
A critical aspect is the initial prompts that guide the GAs' subjective QoE assessment~\cite{liu2023pre}.
The initial prompts setting process is illustrated in Parts A and C of Fig.~\ref{showllm}, including general setup and generative agent-specific settings:
\begin{itemize}
\item In the general setup stage, the Big-Five personality traits are introduced to all $K$ GAs to enhance their understanding and responsiveness to these traits.
\item In the generative agent-specific settings stage, the users' Big-Five personality traits are individually configured for GAs, enabling GAs' similar subjective assessments to given images as real users.
\end{itemize}
With the Big-Five model, the personality of the $k$-${\rm{th}}$ user, i.e., $\bm{u}_k$, can be expressed as
\begin{equation}
\bm{u}_k = \left[ o_k, c_k, e_k, a_k, n_k \right],
\end{equation}
where $k = 1,\ldots,K$, and each element in $\bm{u}_k$ corresponds to a score in one of the Big Five personality traits. Note that the vector $\bm{u}_k$ is significant in our RLLI framework in tailoring the subjective QoE assessment according to individual preferences. Specifically, $\bm{u}_k$ serves as the {\textit{personality traits tuning}} for GAs. Furthermore, $\bm{u}_k$ acts as the user identifier that can be used for the DRL-based resource allocation algorithm design, similar to user representation in AI-based recommendation systems where user preferences are captured and embedded to provide personalized recommendations~\cite{ko2022survey}.

\subsection{Reinforcement Learning With LLMs Feedback Framework}
LLMs are typically designed for language tasks, yet their ability to interpret various task instructions articulated in language has shown promise for acting as universal interfaces for general-purpose assistants~\cite{ma2023llm,ziems2023can}. 
For our proposed RLLI framework, to leverage the inferential capabilities of LLM to simulate users with different personality traits, the LLM-empowered GAs' instructional capacity has to be extended to encompass visual domains.

\subsubsection{Visual Instruction Tuning (VIT)}
{\color{black}The VIT framework extends instruction tuning to process both text and images} using LLMs~\cite{liu2023visual}. With open-source LLMs like LLaMA~\cite{touvron2023llama}, VIT adoption is now more accessible. 
VIT extracts features from images, converting them into language tokens via a trainable projection matrix~\cite{liu2023visual}. Training involves generating multi-turn conversational sequences from images and predicting answer tokens. {\color{black}The framework uses a two-stage instruction tuning process: initially training the visual tokenizer with image-text pairs while keeping the LLM and visual encoder weights fixed, followed by fine-tuning the projection matrix and LLM weights using diverse datasets to enhance response variability. This enables LLM-based GAs to assess image quality.} We use the LLaMA-based LLaVA~\cite{liu2023visual}, an end-to-end trained multi-modal model, for our RLLI algorithm. Note that RLLI's effectiveness is not limited to a specific LLM. Given the generalizable nature of LLMs in understanding and generating language prompts~\cite{liu2023pre,touvron2023llama}, our method is applicable and practical across various LLMs, ensuring broad adaptability and relevance.

\subsubsection{Reinforcement Learning with LLMs Interaction Framework}
RL trains agents to maximize a reward function through interaction with an environment. 
RLHF enhances this process by introducing human insights into the policy optimization~\cite{griffith2013policy}, significantly improving conversational agents like ChatGPT. Nonetheless, both RL and RLHF encounter important challenges:
\begin{itemize}
\item {\textbf{Real-Time Constraint.}} Delayed feedback in RLHF hinder its applicability in scenarios demanding immediate response. While training reward models can mitigate the need for expensive human input, 
	the upfront investment in human data collection and model training can be substantial.
\item {\textbf{Expert Availability.}} Securing consistent access to human experts can be challenging.  Inconsistent interaction and the varying quality of feedback can negatively affect the network management model training.
\item {\textbf{Ethical and Privacy Risk.}} Human-in-the-loop interaction system may present data confidentiality concerns in sensitive applications. For example, some AI-generated images are inappropriate for humans in all ages to view.
\end{itemize}
{\color{black}To address these challenges, we introduce RLLI, where real users can leverage LLM-empowered GAs to provide feedback for DRL model training. 
	These GAs mimic users with varied personalities and provide immediate, context-aware feedback in the form of subjective QoE rewards. 
	Consequently, RLLI offers a real-time, scalable, and financially efficient solution, mitigating the inherent constraints of RL and RLHF. 
	The general algorithm for implementing RLLI is shown as {\textbf{Algorithm}}~\ref{algRLLI}.
	Specifically, the management model initializes with parameters ${\bm{\xi}}$, while $K$ LLM-empowered GAs simulate diverse user feedback. Each episode $e$ begins with state ${\bm{s}}$ and iterates until a terminal state is reached, e.g., task completion, step limit reached, or a failure event.
	Actions ${\bm{a}}$ are generated via policy $\pi_{\bm{\xi}}({\bm{s}})$, with rewards aggregated from the subjective QoE assessments. State transitions and experiences are stored in a replay buffer, facilitating policy parameter updates through experience replay.}

\begin{algorithm}[t]
\caption{Reinforcement Learning with Large Language Model Interaction (RLLI)}
\label{algRLLI}
\textbf{Initialize:} The management model with parameters ${\bm{\xi}}$, LLM-empowered GAs ${\bm K}$ to simulate $K$ users\\
\textbf{Output:} The trained management model ${\bm{\xi}}$
\begin{algorithmic}[1]
\State Input prompts to $K$ GAs, letting them to simulate users with different personalities
\For{each episode \(e = 1, 2, \ldots, E\)}
\State \textbf{Initialize} state \({\bm{s}}\)
\While{\({\bm{s}}\) is not terminal}
\State Generate action \({\bm{a}}\) using policy \(\pi_{\bm{\xi}}({\bm{s}})\)
\State Obtain reward \({{r}} = \sum\limits_{k = 1}^K \text{Agent}_k({\bm{s}}, {\bm{a}}) \)
\State Transition to new state \({\bm{s}}'\)
\State Store transition $({\bm{s}}, {\bm{a}}, r, {\bm{s}}')$ in {\textit{Replay Buffer}}
\State Sample a random minibatch of transitions from {\textit{Replay Buffer}}
\State Update policy parameters ${\bm{\xi}}$
\State \({\bm{s}} \gets {\bm{s}}'\)
\EndWhile
\EndFor
\end{algorithmic}
\end{algorithm}

\subsection{GDM-based DDPG With LLMs Interaction for Joint Resource Allocation}
\begin{figure*}[t]
\centering
\includegraphics[width=0.9\textwidth]{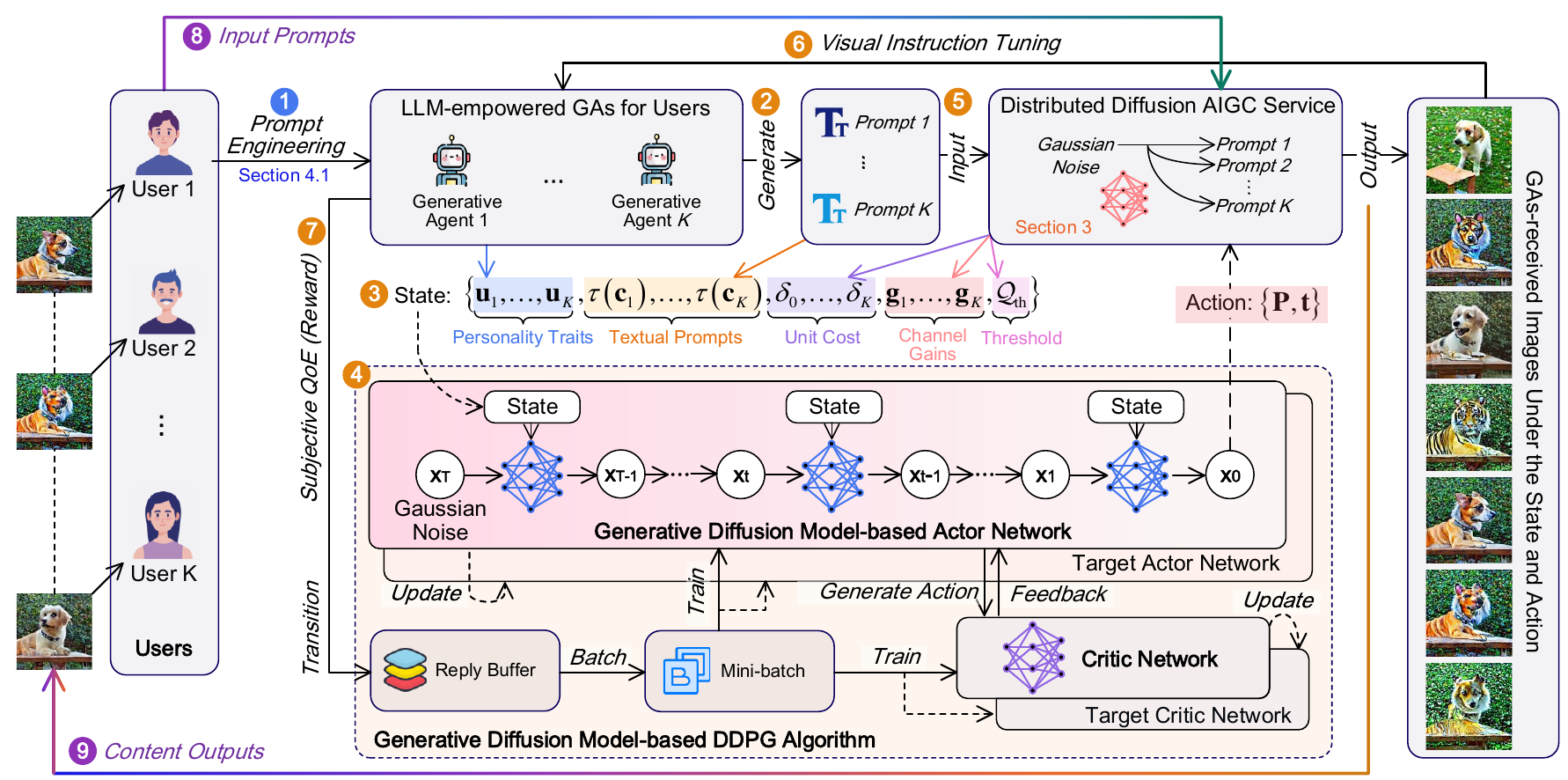}%
\caption{The GDM-aided DDPG with LLMs interaction. Users use prompts to configure LLM-empowered GAs for personality simulation ({\textbf{Step 1}}). In {\textbf{Steps 2 to 7}} within the RLLI framework, GAs generate prompts that, along with channel conditions and other environmental factors, form the {\textit{state}} ({\textbf{Step 3}}). Actions are then generated through a GDM-based actor network ({\textbf{Step 4}}) and images are created via our proposed distributed diffusion AIGC framework. Subsequently, the GAs use these images to generate QoE, providing feedback that serves as a reward for G-DDPG training ({\textbf{Step 7}}), cycling back to {\textbf{Step 2}} for iterative improvement. Upon finalizing the training, users request image generation ({\textbf{Step 8}}), and the trained G-DDPG along with the distributed diffusion AIGC model generates images aimed at maximizing the users' subjective QoE.}
\label{GDDPG}
\end{figure*}
DRL provides an effective solution for resource allocation problems {\color{black}due to its capacity to learn optimal policies in high-dimensional state and action spaces}~\cite{luong2019applications}. Within DRL, DDPG excels in handling continuous action spaces with its actor-critic architecture that ensures stable learning~\cite{lillicrap2015continuous,luong2019applications}. Our problem~\eqref{optimization} involves a hybrid action space of continuous and discrete variables~\cite{9757749}. To manage this, we manipulate the continuous action space to select discrete actions~\cite{9757749,10032267}, such as generating a 10-dimensional action vector with values between 0 and 1 and then choosing the action corresponding to the maximum value. This approach leverages the continuous relaxation of discrete action spaces, enabling gradient-based optimization methods like DDPG to function effectively~\cite{9757749,10032267}. We incorporate a DDPG architecture integrated with a GDM for the actor-network and use a fixed Q-targets strategy to stabilize Q-value updates~\cite{hasselt2010double}. In the following, we explain the action, state, reward, and network training functions of our G-DDPG with the LLMs interaction algorithm.

\textbf{Actions.}
The action is defined as ${\bm{a}}$, which includes the communication and computing resources variables, denoted as ${\bm a} = \left\{ {{\bm{P}},{\bm{t}}} \right\}$, where ${\bm{P}}$ and ${\bm{t}}$ are given as \eqref{dseafa1} and \eqref{dseafa2}, respectively. 
We employ a GDM to generate an optimal communication and computing resource allocation scheme~\cite{du2024diffusion,du2023beyond}, as depicted in Fig.~\ref{GDDPG} and formalized in \eqref{de1}.
The GDM-based method is distinct from the conventional backpropagation or direct optimization approaches typically employed in DRL, offering a progressive refinement of the action distribution.
The action dimensionality, in a scenario where a server caters to $K$ users, is $K + T_{\rm{s}} + K T_{\rm{e}}$. Here, $T_{\rm{s}}$ and $T_{\rm{e}}$ represent the maximum number of denoising steps that can be executed by the edge server and user device, respectively, under the energy constraints.

\textbf{States.} The state vector ${\bm{s}}$ encapsulates variables impacting the optimal communication and computing resource allocation scheme. Thus, ${\bm{s}}$ can be used as the condition in the denoising process, which can be formulated as:
\begin{align}\label{env}
{\bm{s}} = \left\{\!\!\! \begin{array}{l}
{{\bm{u}}_1}, \ldots ,{{\bm{u}}_K},\tau\!\left( {{{\bm{c}}_1}} \right), \ldots ,\tau\!\left( {{{\bm{c}}_K}} \right),{\delta _0}, \ldots ,{\delta _K},\\
{{\bm{g}}_1}, \ldots ,{{\bm{g}}_K},{{\cal Q}_{{\rm{th}}}}
\end{array} \!\!\!\right\},
\end{align}
where ${{\bm{g}}_k}$ $\left( k = 1,\ldots, K\right) $ is the wireless channel gains from the server to the $k$-${\rm{th}}$ device.

\textbf{Reward.}
In designing the reward function for our G-DDPG model, we consider the primary objective of maximizing the sum QoE, i.e., $\sum_{k=1}^{K} \mathcal{Q}_k$, subject to energy constraints, as shown in~\eqref{optimization}. Instead of imposing hard constraints that could lead to sparse rewards and hinder the learning process, we integrate a soft constraint approach into the reward function~\cite{luong2019applications}. This involves subtracting penalty terms for any constraint violations as
\begin{align}\label{fal3eihgfil}
r =& \sum\limits_{k = 1}^K {{{\cal Q}_k}}  - {\lambda _1}\sum\limits_{k = 1}^K {\max \left( {0,{\delta _k}{t_k} - {E_{{T_k}}}} \right)}   \notag\\
& - {\lambda _2}\sum\limits_{k = 1}^K {\max \left( {0,{{\cal Q}_{k}} - {{\cal Q}_{\rm{th}}}} \right)}  \notag\\
& - {\lambda _3}\max \left( {0,{\delta _0}{t_0} + \beta \sum\limits_{k = 1}^K {{P_k}}  - {E_{\rm{T}}}} \right).
\end{align}
The penalties ensure that G-DDPG model learns the significance of constraints while facilitating exploration within the feasible solution space. The coefficients for these penalties can be tuned to balance the trade-off between exploration and adherence to the constraints.

\textbf{Networks.}
The network architecture in G-DDPG comprises actor $A_{\eta}$ and critic $Q_{\upsilon}$ networks, alongside their corresponding target networks $A_{\eta'}$ and $Q_{\upsilon'}$. 
The GDM-based actor network responsible for generating actions and the critic network for evaluating actions. 
The target networks provide tempered, delayed targets, which help in stabilizing the training updates.
The actor network's parameters are updated to maximize the expected reward, while the critic network's parameters are adjusted to minimize the error between the predicted and actual rewards.
Specifically, the training loss for the actor network is given by
\begin{align}\label{xxx2}
\mathop {\arg \min }\limits_{{{A}_{\eta} }} \mathcal{L}_{{A}}({\eta}) = - \mathbb{E}_{{\bm{a}}\sim{{A}_\eta }}\left[ Q_\upsilon \left( {\bm{s}},{\bm{a}} \right) \right].
\end{align}
The critic network loss can be expressed as
\begin{align}\label{xxx1}
\mathop {\arg \min }\limits_{Q_\upsilon } \mathcal{L}_Q(\upsilon) = \mathbb{E}_{{\bm{a}}\sim{{\eta}_\theta }}\left[ \left( Q_\upsilon \left( {\bm{s}},{\bm{a}} \right) - r \right)^2 \right],
\end{align}
where $r$ is the actual Q-value, i.e., reward, obtained from the LLM-empowered GAs after executing the action ${\bm{a}}$. 
The train and inference processes for our proposed G-DDPG with LLMs interaction algorithm is shown in {\textbf{Algorithm~\ref{alg:GDMDDPG_Static}}}. During the training phase, the algorithm iterates over $I$ episodes, each beginning with the GDM-based action generation with the current state as the condition and a noise process for exploration.
{\color{black}The action is executed within the proposed distributed GDM-based AIGC framework, and the resulting reward, derived from GAs' feedback, is stored. A minibatch of experiences is then sampled from the replay buffer to update the networks. Two target networks are updated using a soft update mechanism to ensure stable learning. During inference, the GDM-based actor network generates optimal communication and computing resource allocation strategies based on user prompts.}

\begin{algorithm}[t]
\caption{GDM-aided Double Deep Deterministic Policy Gradient Algorithm}
\label{alg:GDMDDPG_Static}
{\textit{{\textbf{Training Phase:}}}}
\begin{algorithmic}[1]
\State Initialize:
\begin{itemize}
\item Actor network $A_\theta$ and critic network $Q_\upsilon$ with random weights $\theta$ and $\upsilon$, respectively
\item Target networks $A_{\theta'}$ and $Q_{\upsilon'}$ with weights $\theta' \leftarrow \theta$ and $\upsilon' \leftarrow \upsilon$
\item Replay buffer \( \mathcal{B} \)
\end{itemize}
\State Input prompts for $K$ LLM-empowered GAs as shown in Fig.~\ref{showllm}
\For{each iteration $i = 1, 2, \ldots, I$}
\State Initialize a random process \( \mathcal{N} \) for action exploration
\State Observe the current state ${\bm{s}}$
\State {\textbf{Generate action ${\bm{a}}$ according to~\eqref{de1} with ${\bm{s}}$ as the input and exploration noise}}
\State Execute action ${\bm{a}}$ and the distributed GDM-based AIGC framework, and observe reward $r$ from GAs according to~\eqref{fal3eihgfil}
\State Store record $\left({\bm{s}}, {\bm{a}}, r\right)$ in \( \mathcal{B} \)
\State Sample a random minibatch of \( N \) records $\left({\bm{s}}, {\bm{a}}, r\right)$ from \( \mathcal{B} \)
\State Update the critic network~\eqref{xxx1}
\State Update the actor network~\eqref{xxx2}
\State Update the target networks:
\begin{align*}
\theta' \leftarrow \tau \theta + (1 - \tau)\theta' \\
\upsilon' \leftarrow \tau \upsilon + (1 - \tau)\upsilon'
\end{align*}
\State {\textbf{Return}}: The trained GDM-based actor network
\EndFor
\end{algorithmic}
{\textit{{\textbf{Inference Phase:}}}}
\begin{algorithmic}[1]
\State User upload their prompts and form the state ${\bm{s}}$
\State Generate the optimal communication and computing resource allocation scheme ${\bm{a}}$ by using the trained GDM-based actor network
\State {\textbf{Return}}: ${\bm{a}}$
\end{algorithmic}
\end{algorithm}

We then analyze the complexity of {\textbf{Algorithm~\ref{alg:GDMDDPG_Static}}}. Let $w_a$ and $w_c$ denote the actor and critic networks' weight counts, respectively. 
Initialization has a complexity of ${\mathcal{O}}\left(2w_a+2w_c\right)$. The complexity for action generation is augmented to ${\mathcal{O}}(D w_a)$ per iteration due to $D$ denoising steps in the action generation. The replay buffer operations remain ${\mathcal{O}}\left( 1\right) $ for storage and ${\mathcal{O}}\left( N\right) $ for minibatch sampling. 
Updates to the critic and actor networks incur complexities of ${\mathcal{O}}\left(w_c\right) $ and ${\mathcal{O}}\left(w_a\right) $ per update, respectively. 
Target network updates are linear with respect to the number of parameters. Therefore, the training phase now demonstrates an adjusted computational complexity of ${\mathcal{O}}\left(I\left( D w_a + w_c\right) \right)$. 
During the inference phase, generating the optimal resource allocation scheme through the trained actor network necessitates a complexity of ${\mathcal{O}}\left(w_a\right) $, assuming constant-time operations for reward observation and exploration noise generation.

\section{Experiments Results}\label{S5}
The central focus of this paper is the distributed GDM-based AIGC framework and the G-DDPG with LLMs interaction algorithm.
Thus, our experiments are structured to investigate the following questions:
\begin{enumerate}[leftmargin=1.2cm]
\item[{\textbf{Q1)}}] Is the proposed distributed GDM-based AIGC framework effective, with an emphasis on understanding an impact of wireless conditions and diffusion steps on its performance?
\item[{\textbf{Q2)}}] Is the RLLI framework is valid, particularly whether LLM-empowered GAs can mimic human personalities to evaluate images accurately?
\item[{\textbf{Q3)}}] Is the G-DDPG with LLMs interaction algorithm is effective, specifically if it can converge fast and achieve high sum QoE feedback?
\end{enumerate}
We first present the experimental setting and platform, and then answer the above questions through numerical evaluations.

\subsection{Experiments Setting}
We initiated experiments where LLM-empowered GAs are tasked with generating prompts for the training and testing of the distributed GDM-based AIGC framework and the G-DDPG with LLMs interaction algorithm.
Specifically, we considered an environment with one server and three users, i.e., $K=3$. The GAs were first instructed to generate a set of $50$ diverse objectives such as {\textit{``dog''}}, {\textit{``cat''}}, and {\textit{``tree''}}. Subsequently, these GAs combined the objectives to formulate prompts, e.g., {\textit{``dog under the tree''}} and {\textit{``cat on the tree''}}. 
These prompts were then used by the distributed GDM-based AIGC framework and the G-DDPG algorithm for final image generation. 
Then, GAs, each embodying a distinct user personality, provided QoE feedback on image quality, which was used for further G-DDPG model training.
The experimental platform is built on a generic Ubuntu 20.04 system, powered by an AMD Ryzen Threadripper PRO 3975WX 32-Core CPU, and equipped with three NVIDIA RTX A5000 GPUs.

\subsection{Experiments Performance Analysis}
\subsubsection{For {\textbf{Q1}}: Effectiveness of the Distributed GDM-based AIGC Framework}\label{faefg2}
The distributed GDM-based AIGC framework optimizes energy usage by allocating denoising steps among the edge server and user devices while ensuring effective content generation. 
For ease of presentation, we consider two user devices with two different prompts. 
Specifically, {\textit{Device 1}} has {\textit{Prompt 1}}: {\textit{``A tiger sits in front of a bush''}} and {\textit{Device 2}} requires {\textit{Prompt 2}}: {\textit{``A dog sits in front of a bush''}}. 
The server executes $t_0$ denoising steps towards {\textit{Prompt 1}}, and then the two devices separately perform $t_1$ and $t_2$ denoising steps, respectively. 

\begin{figure*}[t]
\centering
\includegraphics[width=0.9\textwidth]{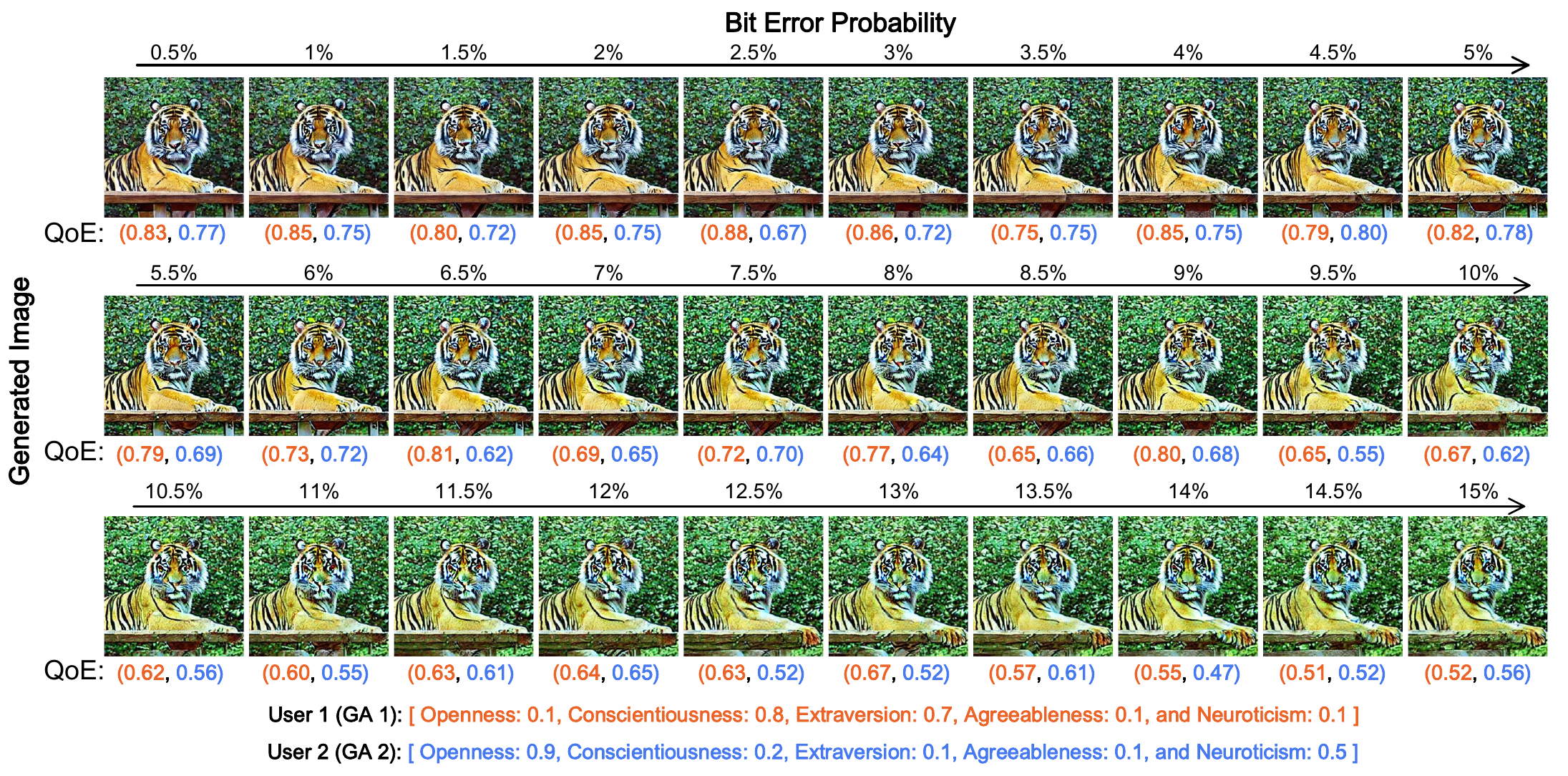}%
\caption{Images generated at {\textit{Device 1}} and corresponding GAs evaluations under varying BEP from $0.5\%$ to $15\%$ within the proposed distributed GDM-based AIGC framework, considering 4 denoising steps at the edge server and 6 denoising steps at {\textit{Device 1}}.}
\label{e1}
\end{figure*}
{\color{black}Diverse power allocation strategies result in different BEPs, which affect the quality of received images. Fig.~\ref{e1} examines the impact of varying BEP levels on image quality at {\textit{Device 1}}. The analysis yields two key observations. First, the user-side subsequent denoising process can compensate for moderate bit errors, indicating that the framework tolerates substantial transmission losses. When the BEP is below $6\%$, the human-perceived image quality at {\textit{Device 1}} remains largely unaffected. 
Second, excessively high BEP degrades the final outcomes because semantic information in intermediate denoised results deviates due to bit errors and cannot be corrected in subsequent steps. This degradation becomes evident when the BEP exceeds $10\%$, where noticeable artifacts appear. Additionally, we employed GAs to simulate users with different personalities and evaluate QoE. The assessment by GAs are consistent with human perception, confirming that the framework has a high tolerance for bit errors within a tolerable threshold.}

\begin{figure*}[t]
\centering
\includegraphics[width=0.9\textwidth]{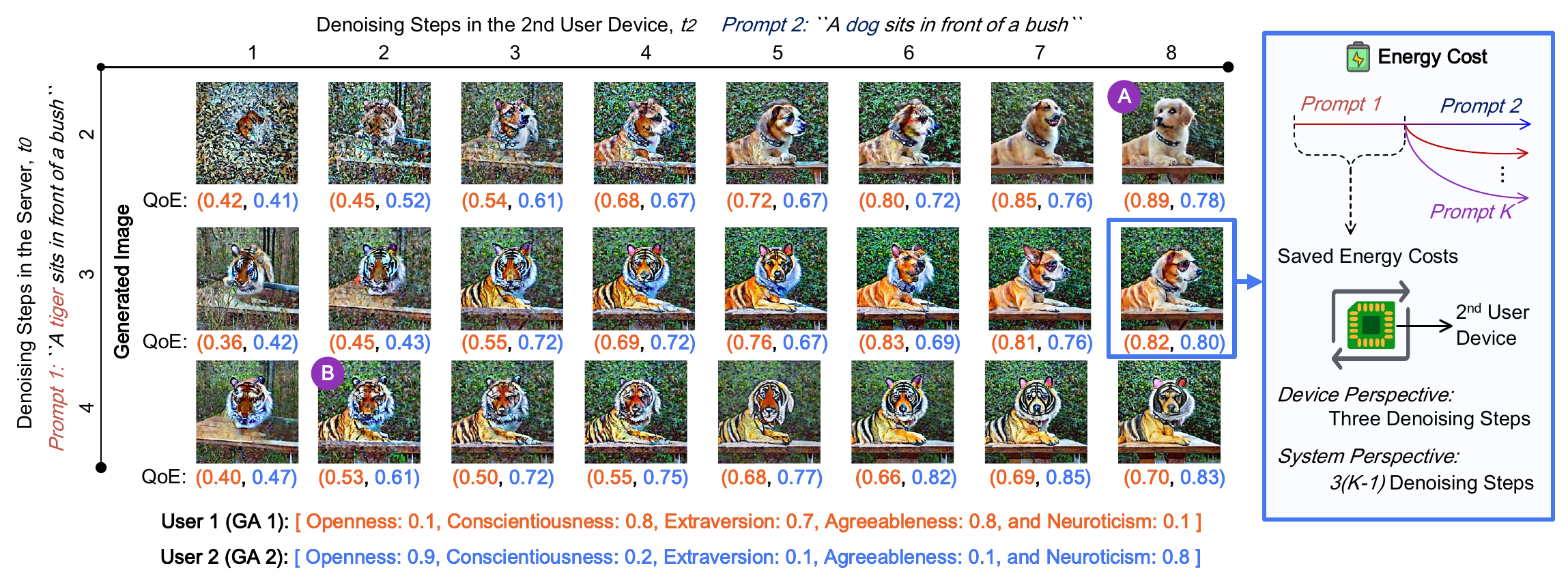}%
\caption{Images generated at {\textit{Device 2}} and corresponding generative agent evaluations under varying the denoising steps at the Server and {\textit{Device 2}} within a distributed GDM-based AIGC framework, considering the BEP is $3\%$.}
\label{e2}
\end{figure*}
The insights from Fig.~\ref{e1} highlight the crucial role of communications resource allocation. Building on this, Fig.~\ref{e2} delves into the computing resource allocation, i.e., the setting of diffusion steps across network devices, and its impact on our distributed GDM-based AIGC framework. 
As shown in Fig.~\ref{e2}, we can observe that an increased number of denoising steps in the server, i.e., $t_0$, enhances the {\textit{``tiger''}} semantics in the intermediate denoised results, influencing {\textit{Device 2}}'s final output.
For example, when $t_0 = 2$ and $t_2 > 4$, the generated dog image on {\textit{Device 2}} barely exhibits tiger characteristics.
However, if $t_0 = 3$, the generated dog on {\textit{Device 2}} is free of tiger features only when $t_2 > 6$. 
With $t_0 = 4$, the generated dog on {\textit{Device 2}} consistently displays tiger-like traits, such as stripes and coloration. 
Additionally, a minimum threshold of computing resources is essential; otherwise, the generated images lack distinctiveness, resulting in poor QoE.
Furthermore, we observe that the semantic fusion of image content has varied effects on users with different personalities. 
For cases where $t_2 = 8$ with different $t_0$ settings, users with high {\textit{agreeableness}} might favor a non-threatening dog. Consequently, the highest QoE is observed when $t_0 = 2$. 
Conversely, users with lower {\textit{agreeableness}} and higher {\textit{openness}} may favor a dog with tiger-like novelty, yielding the highest QoE when $t_0 = 4$.

For energy costs, we consider a general example with a typical smartphone processor in $2$-${\rm{nd}}$ user device, e.g., Qualcomm Snapdragon $870$ processor~\cite{du2023exploring}. As shown in Fig.~\ref{e2}, three steps are reduced in the image generation. If we consider that each step consumes around $1$ to $2$ milliwatt-hours (mWh), i.e., a plausible range for contemporary smartphone CPUs, the removal of three denoising steps could yield energy savings of $3$ to $6$ mWh per execution of the model. Such a reduction would amount to an approximate $20\%$ to $40\%$ decrease in energy consumption for each full cycle of the GDM, relative to the original number of steps. {\color{black}We also observe the balance between QoE requirements and energy cost. When the total energy budget allows for only $8$ denoising steps, no denoising step allocation schemes yield an image rated above a QoE of $0.8$ by any agent. In contrast, with $12$ steps, several allocation schemes result in images rated above $0.8$. This demonstrates that higher energy budgets allow step distributions capable of meeting higher QoE requirements.}
Regarding latency, our distributed architecture ensures minimal delay. Modern network speeds, such as 5G and WiFi 6, facilitate rapid transmission of intermediate latents~\cite{zreikat2020performance}. For example, transmitting a typical 0.1 MB latent in our experiments takes only a few milliseconds, ensuring that the additional network communication time remains negligible.

\subsubsection{For {\textbf{Q2}}: Effectiveness of the RLLI Framework}

\begin{figure*}[t]
	\centering
	\begin{subfigure}[b]{0.24\textwidth}
		\includegraphics[width=\textwidth]{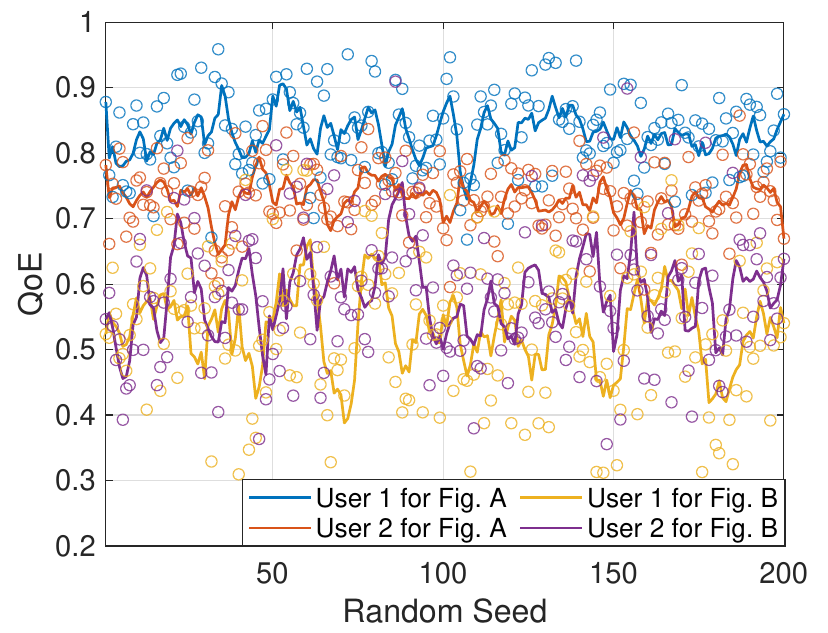}
		\vspace{-0.5cm}
		\caption{}
	\end{subfigure}
	\hfill
	\begin{subfigure}[b]{0.24\textwidth}
		\includegraphics[width=\textwidth]{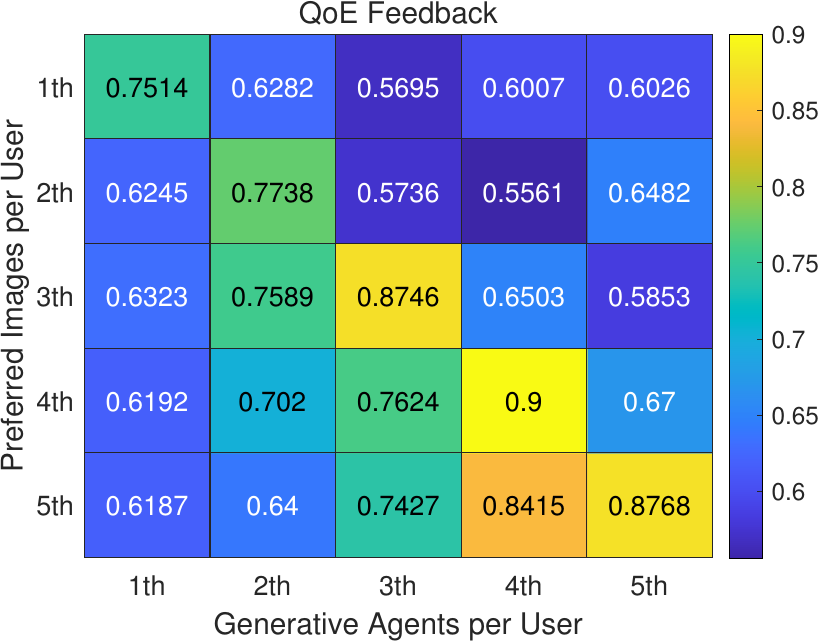}
		\vspace{-0.5cm}
		\caption{}
	\end{subfigure}
	\hfill
	\begin{subfigure}[b]{0.24\textwidth}
		\includegraphics[width=\textwidth]{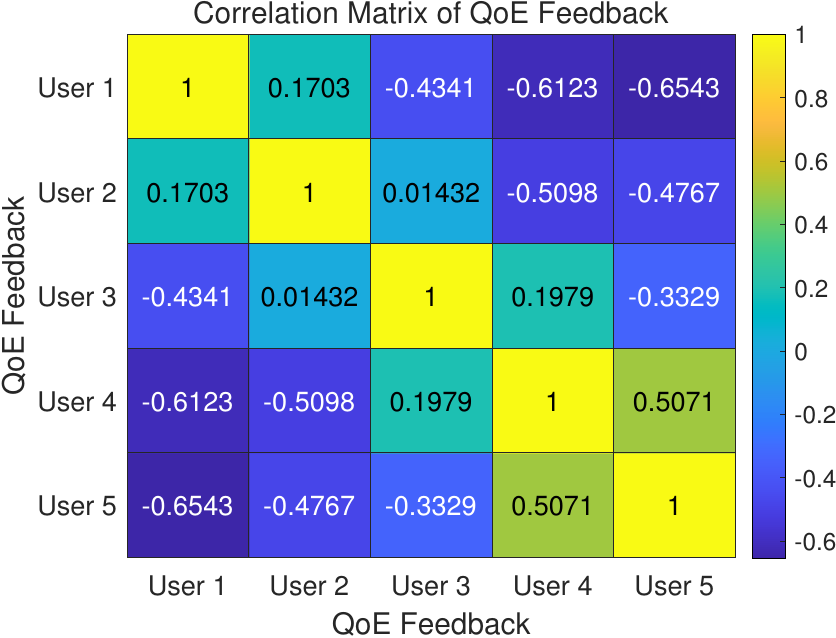}
		\vspace{-0.5cm}
		\caption{}
	\end{subfigure}
	\hfill
	\begin{subfigure}[b]{0.24\textwidth}
		\includegraphics[width=\textwidth]{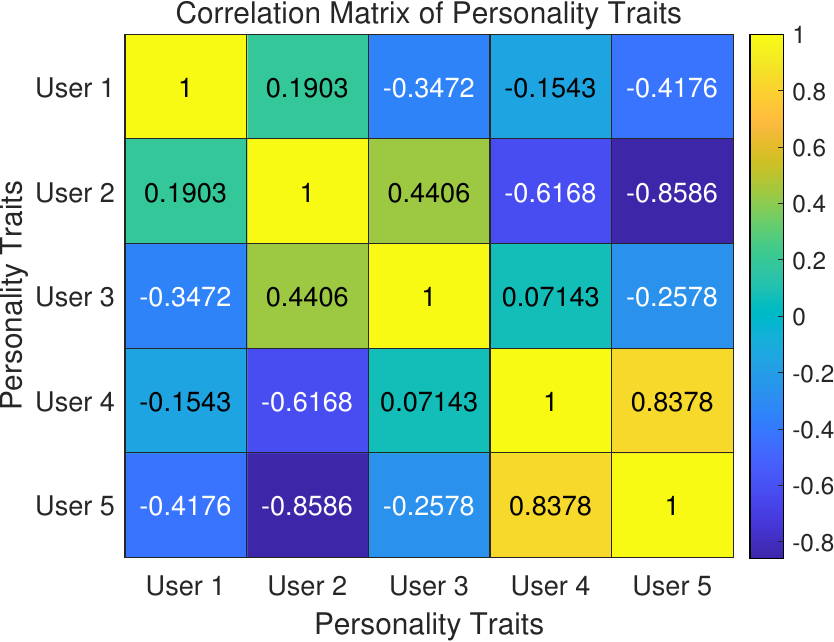}
		\vspace{-0.5cm}
		\caption{}
	\end{subfigure}
\vspace{-0.2cm}
	\caption{(a) The QoE feedback for GAs when they evaluate subfigures A and B of Fig.~\ref{e2}. (b) QoE feedback for five randomly selected users, derived from LLMs simulating their personalities to rate their own and others' preferred images. (c) Correlation matrix of QoE feedback. (d) Correlation matrix of personality traits of five users ([0.65, 0.5, 0.35, 0.75, 0.75], [0.4, 0.8, 0.2, 0.2, 0.8], [0.8, 0.68, 0.56, 0.2, 0.56], [0.8, 0.56, 0.56, 0.68, 0.2], [0.68, 0.56, 0.8, 0.8, 0.2]).}
	\label{EFIG}
\end{figure*}
LLM-empowered GAs have demonstrated their effectiveness in evaluating images, as evidenced in Figs.~\ref{e1} and~\ref{e2}. It is important to note that with a fixed random seed in VIT and LLM setting, GAs can produce the same feedback. Without a fixed seed, a crucial aspect of generative agent feedback for RL is the stability of the evaluation aligned with the personality traits setting of the GAs. 
As shown in Fig.~\ref{EFIG} (a), we use two GAs with different personality settings to evaluate subfigures A and B from Fig.~\ref{e2} across $200$ different random seeds, where the settings of GA1 and GA2 are {\textit{[0.1, 0.8, 0.7, 0.8, 0.1]}} and {\textit{[0.9, 0.2, 0.1, 0.1, 0.5]}}.
We observed that, despite some fluctuations, feedback QoE values clustered for a particular image, suggesting consistency with their personality traits settings. Furthermore, feedback for high-quality images (as in subfigure A) was more stable with less variability, while lower-quality images resulted in greater fluctuations in QoE feedback.

To address the concern regarding the consistency of LLM evaluations with human aesthetic preferences, we conducted a case study using the PsychoFlickr database~\cite{cristani2013unveiling}. Fig.~\ref{EFIG} (b) presents the QoE feedback for five randomly selected users, derived from LLMs simulating their personalities to evaluate both their own preferred images and those of other users. The results show that the LLM assigns the highest scores when evaluating images originally preferred by the users themselves. Figs.~\ref{EFIG} (c) and (d) display the correlation matrices of the QoE feedback and personality traits, respectively. These correlations reveal that the QoE ratings correspond to the correlation of personality traits, whether positive or negative. This finding suggests that the LLM's ratings align well with human preferences when personality traits are similar, thereby providing direct evidence that LLMs can exhibit aesthetic preferences consistent with different personality traits~\cite{jiang2024personallm}.

\subsubsection{For {\textbf{Q3}}: Effectiveness of the G-DDPG with LLMs Interaction Algorithm}
\begin{figure}[!t]
\centering
\includegraphics[width=0.4\textwidth]{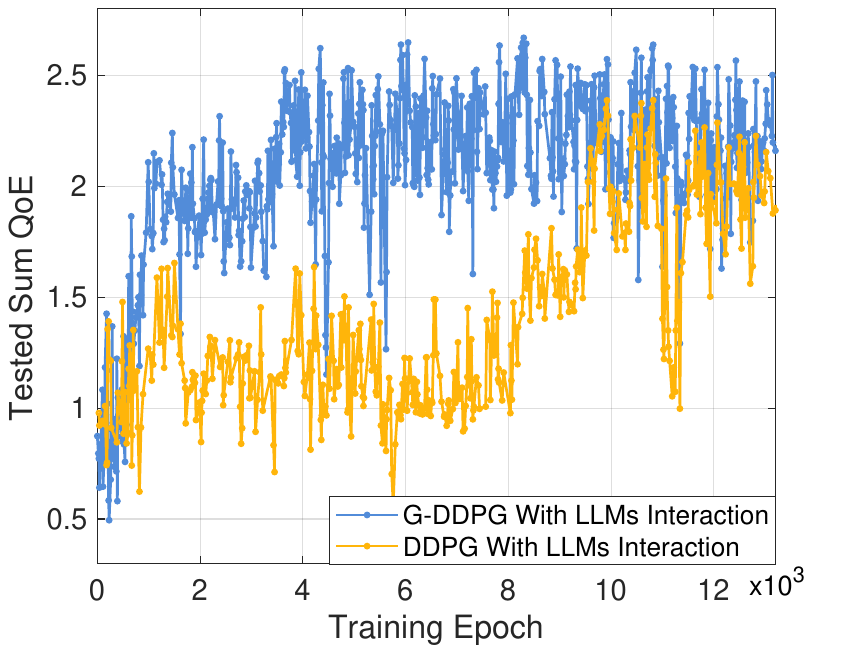}%
\caption{Test reward curves of G-DDPG and DDPG~\cite{lillicrap2015continuous} with LLMs interaction methods, where $K = 3$, $\delta_0$ = $0.5$, $\delta_1 = \delta_2 = \delta_3 = 1$, $E_1 = E_2 = E_3 = 8$, $\mathcal{Q}_{\rm{th}} = 0.6$, $\beta = 0.1$, $E_{\rm{T}} = 20$, learning rate = $10^{-4}$, ${\bm{g}_i} \sim$ Fisher–Snedecor ${\mathcal{F}}$ distribution with the fading severity parameter is $2$ and shadowing amount parameter is $1.5$~\cite{yoo2017fisher}.}
\label{train}
\end{figure}
\begin{figure*}[h]
\centering
\includegraphics[width=0.95\textwidth]{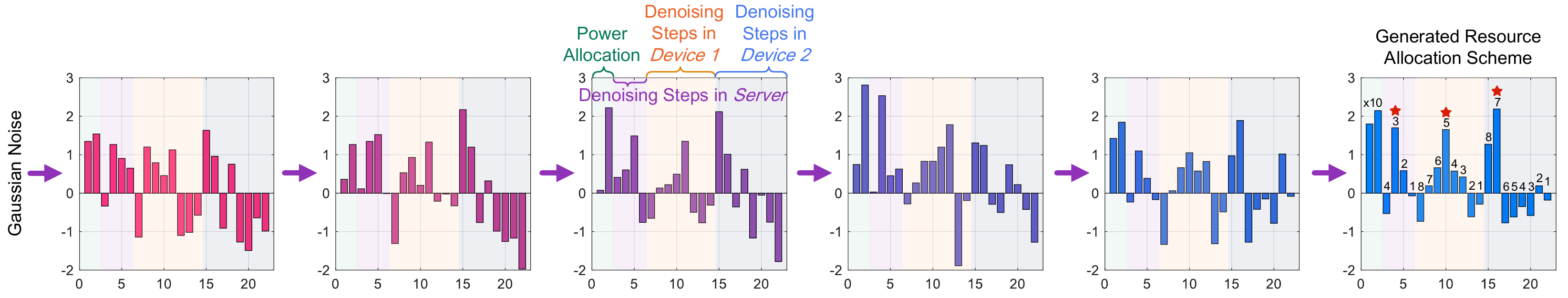}%
\caption{The communication and computing resource allocation scheme in each denoising step of the inference process.}
\label{inference}
\end{figure*}
\begin{figure*}[h]
\centering
\includegraphics[width=0.9\textwidth]{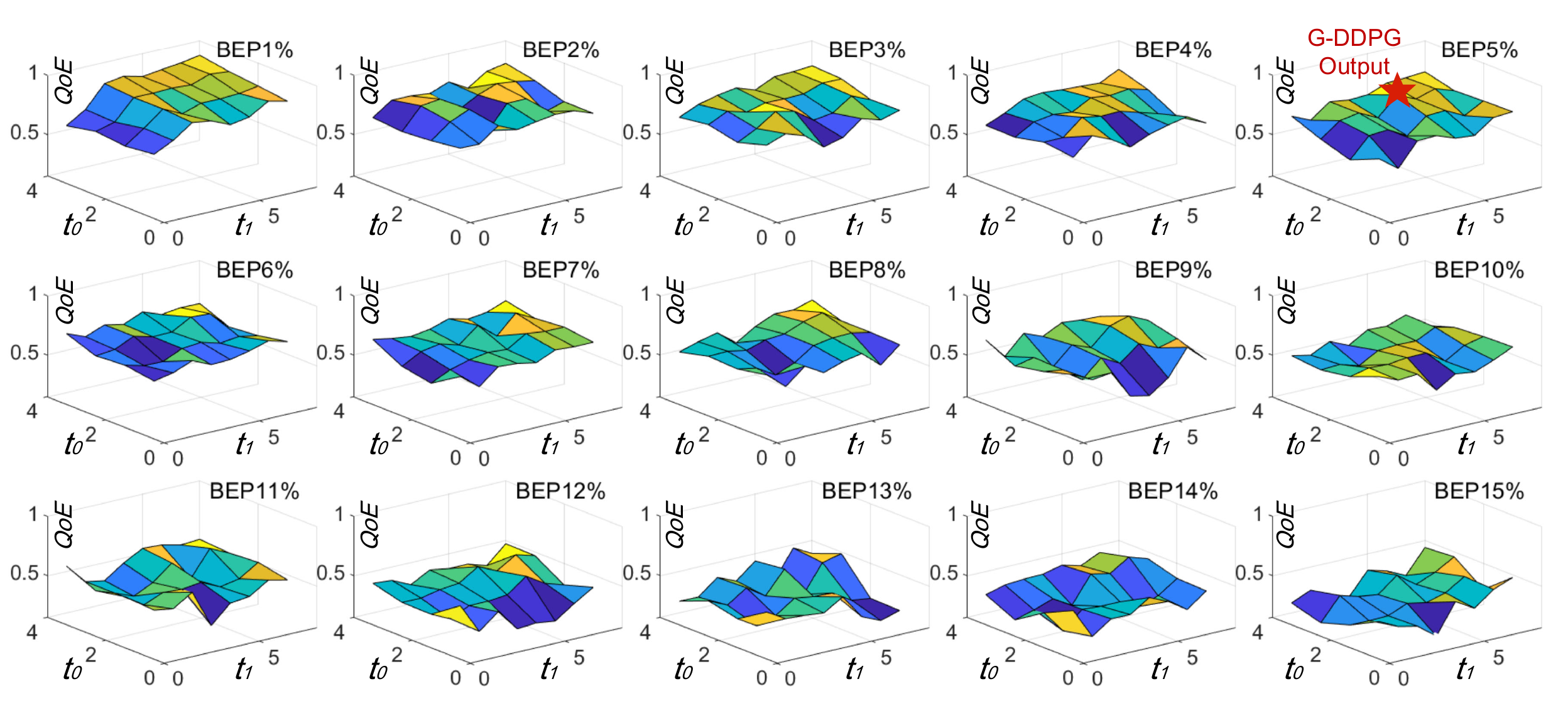}%
\caption{GA's QoE feedback at {\textit{Device 1}} under varying BEP from $0.5\%$ to $15\%$ and different denoising steps within the proposed distributed GDM-based AIGC framework.}
\label{Visio-E31}
\end{figure*}
\begin{figure*}[h]
\centering
\includegraphics[width=0.9\textwidth]{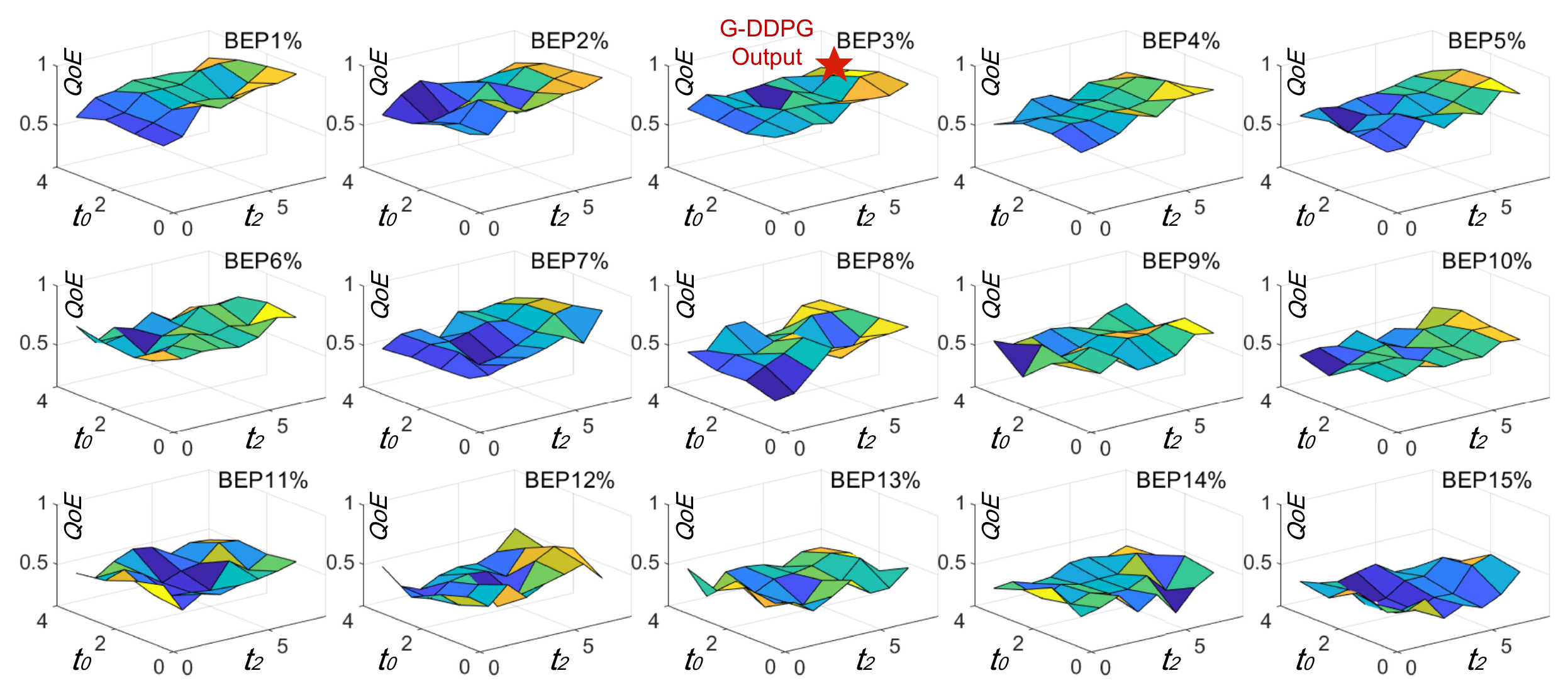}%
\caption{GA's QoE feedback at {\textit{Device 2}} under varying BEP from $0.5\%$ to $15\%$ and different denoising steps within the proposed distributed GDM-based AIGC framework.}
\label{Visio-E32}
\end{figure*}
In Fig.~\ref{train}, we present the training convergence of G-DDPG with LLMs interaction, in comparison with the DDPG algorithm under the RLLI framework. 
While the QoE test curves for both algorithms improve over time, G-DDPG achieves a faster convergence and a higher final sum QoE. 
This superior performance is attributed to the robust modeling capability of the GDM, which facilitates the DDPG algorithm's accelerated learning of the environment and convergence~\cite{du2023beyond}.
Subsequently, we employed the trained model to generate joint communication and computing resource allocation schemes for the case considered in Section~\ref{faefg2}. 
To obtain more insights, Figs.~\ref{Visio-E31} and~\ref{Visio-E32} illustrate the QoE feedback from two GAs across various resource allocation schemes and BEP values. 
A consistent observation is that irrespective of computing resource allocation, insufficient communications resources lead to a higher BEP and result in diminished peak QoE values. 
As shown in Fig.~\ref{Visio-E31}, where both {\textit{Device 1}} and the {\textit{Server}} perform the same prompt, we observe that larger values of $t_0$ and $t_1$, which corresponds to a larger number of total denoising steps towards {\textit{Prompt 1}}, yield higher QoE feedback from the GA. 
However, as shown in Fig.~\ref{Visio-E32}, since {\textit{Device 2}} and the {\textit{Server}} execute different prompts, increasing denoising steps at the {\textit{Server}} does not always enhance QoE, as semantic integration may compromise image quality to some extent. In this case, peak QoE values are achieved when $t_0$ is moderate, and $t_2$ is sufficient.
{\color{black}We further examine how QoE responds to different allocation schemes under a fixed energy budget $E_{\rm{T}}$. In subfigures with low BEP (e.g., more energy for communication), QoE is limited when $t_0$ is small due to insufficient computation. In contrast, in high-BEP subfigures (e.g., less energy for communication), QoE also drops at small $t_0$ due to degraded transmission quality. These variations show that even under the same budget, allocation schemes significantly affect QoE, underscoring the need for careful optimization. Furthermore, the red pentagram marks the joint communication and computing resource allocation scheme under G-DDPM.} 
As shown in Fig.~\ref{inference}, we obtain $t_0 = 3$, $t_1 = 5$, $t_2 = 7$, and the transmission power from the {\textit{Server}} to {\textit{Device 1}} and {\textit{Device 2}} are $18$ and $21$ ${\rm{dBW}}$, respectively. This leads that {\textit{Device 1}}'s BEP being $5\%$ and {\textit{Device 2}}'s BEP being $3\%$, as shown in Figs.~\ref{Visio-E31} and~\ref{Visio-E32}. Besides increased denoising steps, more communications resources are allocated to {\textit{Device 2}} to minimize bit error in the interrupted results, thus reducing the complexity of directional semantic changes in the denoising process. Consequently, the maximum total QoE is obtained.

\section{Conclusion}\label{S6}
We have introduced a distributed GDM-based AIGC framework that {\color{black}improves} energy efficiency and maximize the subjective QoE {\color{black}through} the proposed IAI method, i.e., RLLI algorithm. 
{\color{black}By restructuring the denoising process in GDMs, the framework reduces the total number of denoising steps by enabling shared diffusion for semantically similar prompts, thereby conserving resources while meeting user requirements.}
The proposed RLLI algorithm is an interactive AI method, which utilizes LLM-empowered GAs to provide subjective and real-time QoE feedback for AI-based resource allocation model training. 
Specifically, {\color{black}the} proposed G-DDPG with LLMs interaction algorithm optimizes communication and computing resource allocation {\color{black}by accounting for user personalities and dynamic wireless conditions. Simulations demonstrated a substantial improvement in the sum QoE, up to $15\%$, compared with conventional DDPG methods.}
Future research directions include:
\begin{itemize}
\item {\color{black}{\textit{Caching Mechanism.}} Investigating a caching mechanism that stores intermediate denoised results for reuse, allowing the server to return cached outputs when new user prompts are similar to previous ones. This reduces redundant computation and mitigates the need for immediate processing of each query. Such a mechanism also relaxes the requirement for real-time QoE feedback of LLM agents by decoupling response generation from on-demand model inference. A central challenge is to balance cache storage and access strategies against the achievable energy and latency savings.}
\item {\textit{Dataset for GAs.}} Establishing a dataset through surveys with natural human users to improve the simulation of user preferences by LLM-empowered GAs, thereby enhancing the accuracy of subjective QoE assessments.
\item {\color{black} {\textit{Optimized Computation Reuse via LLMs.}} Leveraging LLMs to optimize computational reuse during inference by clustering semantically similar user prompts and integrating GDM characteristics into the reasoning process. This strategy can substantially reduce computational overhead without compromising output quality. Additionally, developing simulation-based evaluation methodologies is essential to assess computational efficiency and generation quality within realistic multi-model collaborative environments.}
\end{itemize}

\bibliographystyle{IEEEtran}
\bibliography{Ref}

\begin{thebibliography}{10}
\providecommand{\url}[1]{#1}
\csname url@samestyle\endcsname
\providecommand{\newblock}{\relax}
\providecommand{\bibinfo}[2]{#2}
\providecommand{\BIBentrySTDinterwordspacing}{\spaceskip=0pt\relax}
\providecommand{\BIBentryALTinterwordstretchfactor}{4}
\providecommand{\BIBentryALTinterwordspacing}{\spaceskip=\fontdimen2\font plus
\BIBentryALTinterwordstretchfactor\fontdimen3\font minus
  \fontdimen4\font\relax}
\providecommand{\BIBforeignlanguage}[2]{{%
\expandafter\ifx\csname l@#1\endcsname\relax
\typeout{** WARNING: IEEEtran.bst: No hyphenation pattern has been}%
\typeout{** loaded for the language `#1'. Using the pattern for}%
\typeout{** the default language instead.}%
\else
\language=\csname l@#1\endcsname
\fi
#2}}
\providecommand{\BIBdecl}{\relax}
\BIBdecl

\bibitem{guo2022systematic}
X.~Guo and L.~Zhao, ``A systematic survey on deep generative models for graph
  generation,'' \emph{IEEE Trans. Pattern Anal. Mach. Intell.}, vol.~45, no.~5,
  pp. 5370--5390, May 2022.

\bibitem{du2023enabling}
H.~Du, Z.~Li, D.~Niyato, J.~Kang, Z.~Xiong, D.~I. Kim \emph{et~al.}, ``Enabling
  {AI}-generated content ({AIGC}) services in wireless edge networks,''
  \emph{IEEE Wireless Mag.}, to appear, 2023.

\bibitem{lund2023chatting}
B.~D. Lund and T.~Wang, ``Chatting about {ChatGPT}: how may {AI} and {GPT}
  impact academia and libraries?'' \emph{Library Hi Tech News}, vol.~40, no.~3,
  pp. 26--29, Mar. 2023.

\bibitem{stabdiff}
S.~AI, ``Stable diffusion,'' \url{https://stability.ai/}, access time: Oct.
  10th, 2025.

\bibitem{croitoru2023diffusion}
F.-A. Croitoru, V.~Hondru, R.~T. Ionescu, and M.~Shah, ``Diffusion models in
  vision: {A} survey,'' \emph{IEEE Trans. Pattern Anal. Mach. Intell.}, to
  appear, 2023.

\bibitem{amershi2019guidelines}
S.~Amershi, D.~Weld, M.~Vorvoreanu, A.~Fourney, B.~Nushi, P.~Collisson, J.~Suh,
  S.~Iqbal, P.~N. Bennett, K.~Inkpen \emph{et~al.}, ``Guidelines for human-{AI}
  interaction,'' in \emph{Proc. 2019 Chi Conf. Human Fact. Comput. Syst.},
  2019, pp. 1--13.

\bibitem{picard2001toward}
R.~W. Picard, E.~Vyzas, and J.~Healey, ``Toward machine emotional intelligence:
  Analysis of affective physiological state,'' \emph{IEEE Trans. Pattern Anal.
  Mach. Intell.}, vol.~23, no.~10, pp. 1175--1191, Oct. 2001.

\bibitem{juluri2015measurement}
P.~Juluri, V.~Tamarapalli, and D.~Medhi, ``Measurement of quality of experience
  of video-on-demand services: {A} survey,'' \emph{IEEE Commun. Surv. Tut.},
  vol.~18, no.~1, pp. 401--418, Jan. 2015.

\bibitem{du2023attention}
H.~Du, J.~Liu, D.~Niyato, J.~Kang, Z.~Xiong, J.~Zhang, and D.~I. Kim,
  ``Attention-aware resource allocation and {QoE} analysis for metaverse
  {xURLLC} services,'' \emph{IEEE J. Selec. Areas Commun.}, vol.~41, no.~7,
  Jul. 2023.

\bibitem{luong2019applications}
N.~C. Luong, D.~T. Hoang, S.~Gong, D.~Niyato, P.~Wang, Y.-C. Liang, and D.~I.
  Kim, ``Applications of deep reinforcement learning in communications and
  networking: {A} survey,'' \emph{IEEE Commun. Surv. Tut.}, vol.~21, no.~4, pp.
  3133--3174, Apr. 2019.

\bibitem{du2023beyond}
H.~Du, R.~Zhang, Y.~Liu, J.~Wang, Y.~Lin, Z.~Li, D.~Niyato, J.~Kang, Z.~Xiong,
  S.~Cui \emph{et~al.}, ``Beyond deep reinforcement learning: {A} tutorial on
  generative diffusion models in network optimization,'' \emph{{\rm{arXiv
  preprint arXiv:2308.05384}}}, 2023.

\bibitem{liu2023pre}
P.~Liu, W.~Yuan, J.~Fu, Z.~Jiang, H.~Hayashi, and G.~Neubig, ``Pre-train,
  prompt, and predict: {A} systematic survey of prompting methods in natural
  language processing,'' \emph{ACM Comput. Surv.}, vol.~55, no.~9, pp. 1--35,
  Sep. 2023.

\bibitem{luo2023latent}
S.~Luo, Y.~Tan, L.~Huang, J.~Li, and H.~Zhao, ``Latent consistency models:
  Synthesizing high-resolution images with few-step inference,''
  \emph{{\rm{arXiv preprint arXiv:2310.04378}}}, 2023.

\bibitem{sauer2023adversarial}
A.~Sauer, D.~Lorenz, A.~Blattmann, and R.~Rombach, ``Adversarial diffusion
  distillation,'' \emph{{\rm{arXiv preprint arXiv:2311.17042}}}, 2023.

\bibitem{geng2023visual}
D.~Geng, I.~Park, and A.~Owens, ``Visual anagrams: Generating multi-view
  optical illusions with diffusion models,'' \emph{{\rm{arXiv preprint
  arXiv:2311.17919}}}, 2023.

\bibitem{du2023demofusion}
R.~Du, D.~Chang, T.~Hospedales, Y.-Z. Song, and Z.~Ma, ``Demo{F}usion:
  Democratising high-resolution image generation with no \$\$\$,''
  \emph{{\rm{arXiv preprint arXiv:2311.16973}}}, 2023.

\bibitem{sora}
O.~AI, ``Sora,'' \url{https://openai.com/index/sora/}.

\bibitem{xu2023unleashing}
M.~Xu, H.~Du, D.~Niyato, J.~Kang, Z.~Xiong, S.~Mao, Z.~Han, A.~Jamalipour,
  D.~I. Kim, V.~Leung \emph{et~al.}, ``Unleashing the power of edge-cloud
  generative {AI} in mobile networks: {A} survey of {AIGC} services,''
  \emph{IEEE Commun. Surv. Tutorials}, vol.~26, no.~2, pp. 1127--1170, Jan.
  2024.

\bibitem{dang2020should}
S.~Dang, O.~Amin, B.~Shihada, and M.-S. Alouini, ``What should {6G} be?''
  \emph{Nat. Electron.}, vol.~3, no.~1, pp. 20--29, Jan. 2020.

\bibitem{ma2023llm}
X.~Ma, G.~Fang, and X.~Wang, ``{LLM}-{P}runer: {O}n the structural pruning of
  large language models,'' \emph{{\rm{arXiv preprint arXiv:2305.11627}}}, 2023.

\bibitem{kim2023architectural}
B.-K. Kim, H.-K. Song, T.~Castells, and S.~Choi, ``On architectural compression
  of text-to-image diffusion models,'' \emph{{\rm{arXiv preprint
  arXiv:2305.15798}}}, 2023.

\bibitem{du2023exploring}
H.~Du, R.~Zhang, D.~Niyato, J.~Kang, Z.~Xiong, D.~I. Kim, X.~S. Shen, and H.~V.
  Poor, ``Exploring collaborative distributed diffusion-based {AI}-generated
  content ({AIGC}) in wireless networks,'' \emph{IEEE Netw.}, no.~99, pp. 1--8,
  2023.

\bibitem{ho2020denoising}
J.~Ho, A.~Jain, and P.~Abbeel, ``Denoising diffusion probabilistic models,''
  \emph{Adv. Neural Inf. Process. Syst.}, vol.~33, pp. 6840--6851, 2020.

\bibitem{ho2022video}
J.~Ho, T.~Salimans, A.~Gritsenko, W.~Chan, M.~Norouzi, and D.~J. Fleet, ``Video
  diffusion models,'' \emph{{\rm{arXiv:2204.03458}}}, 2022.

\bibitem{li2022diffusion}
X.~Li, J.~Thickstun, I.~Gulrajani, P.~S. Liang, and T.~B. Hashimoto,
  ``Diffusion-lm improves controllable text generation,'' \emph{Adv. Neural
  Inf. Process. Syst.}, vol.~35, pp. 4328--4343, 2022.

\bibitem{reid2023diffuser}
M.~Reid, V.~J. Hellendoorn, and G.~Neubig, ``Diffuser: {D}iffusion via
  edit-based reconstruction,'' \emph{Proc. Int. Conf. Learn. Represent.}, 2023.

\bibitem{huang2022prodiff}
R.~Huang, Z.~Zhao, H.~Liu, J.~Liu, C.~Cui, and Y.~Ren, ``Prodiff: {P}rogressive
  fast diffusion model for high-quality text-to-speech,'' in \emph{Proc. ACM
  Int. Conf. Multimedia}, 2022, pp. 2595--2605.

\bibitem{kongdiffwave}
Z.~Kong, W.~Ping, J.~Huang, K.~Zhao, and B.~Catanzaro, ``Diff{W}ave: {A}
  versatile diffusion model for audio synthesis,'' in \emph{Proc. Int. Conf.
  Learn. Represent.}, 2021.

\bibitem{niu2020permutation}
C.~Niu, Y.~Song, J.~Song, S.~Zhao, A.~Grover, and S.~Ermon, ``Permutation
  invariant graph generation via score-based generative modeling,'' in
  \emph{Proc. Int. Conf. Artif. Intell. Stat.}, 2020, pp. 4474--4484.

\bibitem{ketata2023diffdock}
M.~A. Ketata, C.~Laue, R.~Mammadov, H.~Stark, M.~Wu, G.~Corso, C.~Marquet,
  R.~Barzilay, and T.~S. Jaakkola, ``Diff{D}ock-{PP}: {R}igid protein-protein
  docking with diffusion models,'' in \emph{Proc. Int. Conf. Learn.
  Represent.}, 2023.

\bibitem{li2020personality}
L.~Li, H.~Zhu, S.~Zhao, G.~Ding, and W.~Lin, ``Personality-assisted multi-task
  learning for generic and personalized image aesthetics assessment,''
  \emph{IEEE Trans. Image Process.}, vol.~29, pp. 3898--3910, Jan. 2020.

\bibitem{saucier1998beyond}
G.~Saucier and L.~R. Goldberg, ``What is beyond the {B}ig {F}ive?'' \emph{J.
  Pers.}, vol.~66, pp. 495--524, 1998.

\bibitem{cristani2013unveiling}
M.~Cristani, A.~Vinciarelli, C.~Segalin, and A.~Perina, ``Unveiling the
  multimedia unconscious: {I}mplicit cognitive processes and multimedia content
  analysis,'' in \emph{Proc. ACM Int. Conf. Multimedia}, 2013, pp. 213--222.

\bibitem{wang2023chatgpt}
J.~Wang, Y.~Liang, F.~Meng, H.~Shi, Z.~Li, J.~Xu, J.~Qu, and J.~Zhou, ``Is
  {C}hat{GPT} a good {NLG} evaluator? {A} preliminary study,'' \emph{{\rm{arXiv
  preprint arXiv:2303.04048}}}, 2023.

\bibitem{lee2023rlaif}
H.~Lee, S.~Phatale, H.~Mansoor, K.~Lu, T.~Mesnard, C.~Bishop, V.~Carbune, and
  A.~Rastogi, ``Rlaif: {S}caling reinforcement learning from human feedback
  with {AI} feedback,'' \emph{{\rm{arXiv preprint arXiv:2309.00267}}}, 2023.

\bibitem{ziems2023can}
C.~Ziems, W.~Held, O.~Shaikh, J.~Chen, Z.~Zhang, and D.~Yang, ``Can large
  language models transform computational social science?'' \emph{{\rm{arXiv
  preprint arXiv:2305.03514}}}, 2023.

\bibitem{wang2023does}
X.~Wang, Y.~Fei, Z.~Leng, and C.~Li, ``Does role-playing chatbots capture the
  character personalities? {A}ssessing personality traits for role-playing
  chatbots,'' \emph{arXiv preprint arXiv:2310.17976}, 2023.

\bibitem{sohl2015deep}
J.~Sohl-Dickstein, E.~Weiss, N.~Maheswaranathan, and S.~Ganguli, ``Deep
  unsupervised learning using nonequilibrium thermodynamics,'' in \emph{Proc.
  Int. Conf. Mach. Learn.}\hskip 1em plus 0.5em minus 0.4em\relax PMLR, 2015,
  pp. 2256--2265.

\bibitem{song2019generative}
Y.~Song and S.~Ermon, ``Generative modeling by estimating gradients of the data
  distribution,'' \emph{Adv. Neural Inf. Process. Syst.}, vol.~32, 2019.

\bibitem{song2020score}
Y.~Song, J.~Sohl-Dickstein, D.~P. Kingma, A.~Kumar, S.~Ermon, and B.~Poole,
  ``Score-based generative modeling through stochastic differential
  equations,'' \emph{{\rm{arXiv preprint arXiv:2011.13456}}}, 2020.

\bibitem{kingma2021variational}
D.~Kingma, T.~Salimans, B.~Poole, and J.~Ho, ``Variational diffusion models,''
  \emph{Adv. Neural Inf. Process. Syst.}, vol.~34, pp. 21\,696--21\,707, 2021.

\bibitem{ronneberger2015u}
O.~Ronneberger, P.~Fischer, and T.~Brox, ``U-net: {C}onvolutional networks for
  biomedical image segmentation,'' in \emph{In MICCAI (3), volume 9351 of
  Lecture Notes in Computer Science}.\hskip 1em plus 0.5em minus 0.4em\relax
  Springer, 2015, pp. 234--241.

\bibitem{gauthier2014conditional}
J.~Gauthier, ``Conditional generative adversarial nets for convolutional face
  generation,'' \emph{Class project for Stanford CS231N: convolutional neural
  networks for visual recognition, Winter semester}, vol. 2014, no.~5, p.~2,
  2014.

\bibitem{sohn2015learning}
K.~Sohn, H.~Lee, and X.~Yan, ``Learning structured output representation using
  deep conditional generative models,'' \emph{Adv. Neural Inf. Process. Syst.},
  vol.~28, 2015.

\bibitem{reed2016generative}
S.~Reed, Z.~Akata, X.~Yan, L.~Logeswaran, B.~Schiele, and H.~Lee, ``Generative
  adversarial text to image synthesis,'' in \emph{Proc. Int. Conf. Mach.
  Learn.}\hskip 1em plus 0.5em minus 0.4em\relax PMLR, 2016, pp. 1060--1069.

\bibitem{park2019semantic}
T.~Park, M.-Y. Liu, T.-C. Wang, and J.-Y. Zhu, ``Semantic image synthesis with
  spatially-adaptive normalization,'' in \emph{Proc. IEEE Conf. Comput. Vis.
  Pattern Recognit.}, 2019, pp. 2337--2346.

\bibitem{isola2017image}
P.~Isola, J.-Y. Zhu, T.~Zhou, and A.~A. Efros, ``Image-to-image translation
  with conditional adversarial networks,'' in \emph{Proc. IEEE Conf. Comput.
  Vis. Pattern Recognit.}, 2017, pp. 1125--1134.

\bibitem{ko2022survey}
H.~Ko, S.~Lee, Y.~Park, and A.~Choi, ``A survey of recommendation systems:
  {R}ecommendation models, techniques, and application fields,''
  \emph{Electronics}, vol.~11, no.~1, p. 141, Jan. 2022.

\bibitem{liu2023visual}
H.~Liu, C.~Li, Q.~Wu, and Y.~J. Lee, ``Visual instruction tuning,''
  \emph{{\rm{arXiv preprint arXiv:2304.08485}}}, 2023.

\bibitem{touvron2023llama}
H.~Touvron, T.~Lavril, G.~Izacard, X.~Martinet, M.-A. Lachaux, T.~Lacroix,
  B.~Rozi{\`e}re, N.~Goyal, E.~Hambro, F.~Azhar \emph{et~al.}, ``Llama: {O}pen
  and efficient foundation language models,'' \emph{{\rm{arXiv preprint
  arXiv:2302.13971}}}, 2023.

\bibitem{griffith2013policy}
S.~Griffith, K.~Subramanian, J.~Scholz, C.~L. Isbell, and A.~L. Thomaz,
  ``Policy shaping: {I}ntegrating human feedback with reinforcement learning,''
  \emph{Adv. Neural Inf. Process. Syst.}, vol.~26, 2013.

\bibitem{lillicrap2015continuous}
T.~P. Lillicrap, J.~J. Hunt, A.~Pritzel, N.~Heess, T.~Erez, Y.~Tassa,
  D.~Silver, and D.~Wierstra, ``Continuous control with deep reinforcement
  learning,'' \emph{{\rm{arXiv preprint arXiv:1509.02971}}}, 2015.

\bibitem{9757749}
L.~Ale, S.~A. King, N.~Zhang, A.~R. Sattar, and J.~Skandaraniyam, ``{D3PG}:
  Dirichlet {DDPG} for task partitioning and offloading with constrained hybrid
  action space in mobile-edge computing,'' \emph{IEEE Internet Things J.},
  vol.~9, no.~19, pp. 19\,260--19\,272, 2022.

\bibitem{10032267}
R.~Zhang, K.~Xiong, Y.~Lu, P.~Fan, D.~W.~K. Ng, and K.~B. Letaief, ``Energy
  efficiency maximization in {RIS}-assisted {SWIPT} networks with {RSMA}: {A}
  {PPO}-based approach,'' \emph{IEEE J. Sel. Areas Commun.}, vol.~41, no.~5,
  pp. 1413--1430, May 2023.

\bibitem{hasselt2010double}
H.~Hasselt, ``Double {Q}-learning,'' \emph{Adv. Neural Inf. Process. Syst.},
  vol.~23, 2010.

\bibitem{du2024diffusion}
H.~Du, Z.~Li, D.~Niyato, J.~Kang, Z.~Xiong, H.~Huang, and S.~Mao,
  ``Diffusion-based reinforcement learning for edge-enabled {AI}-generated
  content services,'' \emph{IEEE Trans. Mobile Comput.}, 2024.

\bibitem{zreikat2020performance}
E.~J. Oughton, W.~Lehr, K.~Katsaros, I.~Selinis, D.~Bubley, and J.~Kusuma,
  ``Revisiting wireless internet connectivity: {5G} vs {Wi-Fi} 6,''
  \emph{Telecommun. Policy}, vol.~45, no.~5, p. 102127, May 2021.

\bibitem{jiang2024personallm}
H.~Jiang, X.~Zhang, X.~Cao, C.~Breazeal, D.~Roy, and J.~Kabbara,
  ``{PersonaLLM}: Investigating the ability of large language models to express
  personality traits,'' in \emph{Findings of the Association for Computational
  Linguistics: NAACL 2024}, 2024, pp. 3605--3627.

\bibitem{yoo2017fisher}
S.~K. Yoo, S.~L. Cotton, P.~C. Sofotasios, M.~Matthaiou, M.~Valkama, and G.~K.
  Karagiannidis, ``The {F}isher-{S}nedecor ${\mathcal{f}}$ distribution: {A}
  simple and accurate composite fading model,'' \emph{IEEE Commun. Lett.},
  vol.~21, no.~7, pp. 1661--1664, Jul. 2017.

\end{thebibliography}

\begin{IEEEbiography}[{\includegraphics[width=1in,height=1.25in, clip,keepaspectratio]{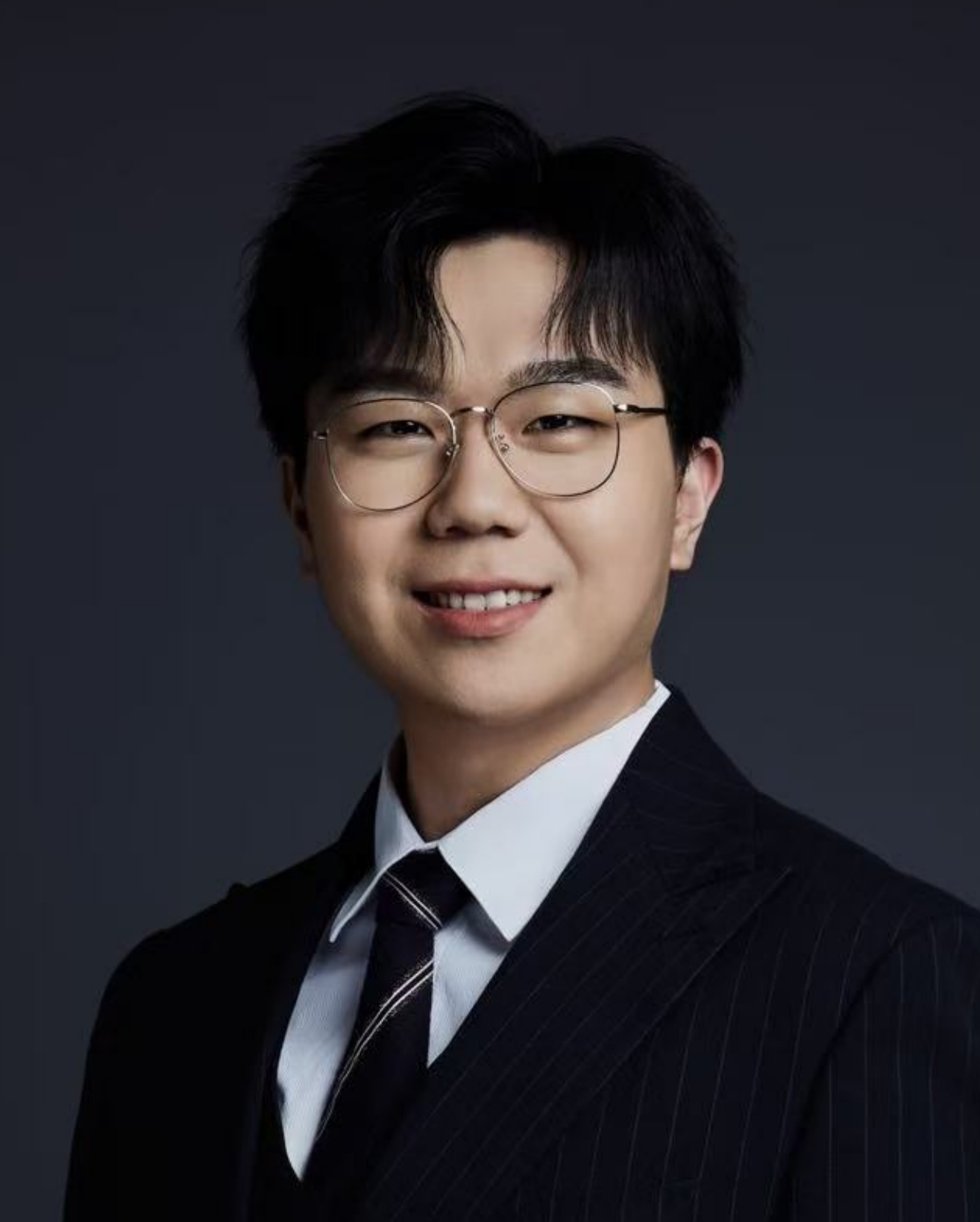}}]{Hongyang Du} is an assistant professor at the Department of Electrical and Electronic Engineering, The University of Hong Kong. He received the BEng degree from Beijing Jiaotong University, Beijing, and the PhD degree from the Nanyang Technological University, Singapore. He serves as the Editor-in-Chief Assistant (2022-2024) and Editor (2025-Present) of IEEE Communications Surveys \& Tutorials, the Editor of IEEE Transactions on Communications, IEEE Transactions on Vehicular Technology, IEEE Open Journal of the Communications Society, and the Guest Editor for IEEE Vehicular Technology Magazine. He is the recipient of the IEEE ComSoc Young Professional Award for Best Early Career Researcher in 2024, IEEE Daniel E. Noble Fellowship Award from the IEEE Vehicular Technology Society in 2022, the IEEE Signal Processing Society Scholarship from the IEEE Signal Processing Society in 2023, and the Singapore Data Science Consortium (SDSC) Dissertation Research Fellowship in 2023. He was recognized as an exemplary reviewer of the IEEE Transactions on Communications and IEEE Communications Letters. His research interests include edge intelligence, generative AI, and network management.
\end{IEEEbiography}

\begin{IEEEbiography}[{\includegraphics[width=1in,height=1.25in, clip,keepaspectratio]{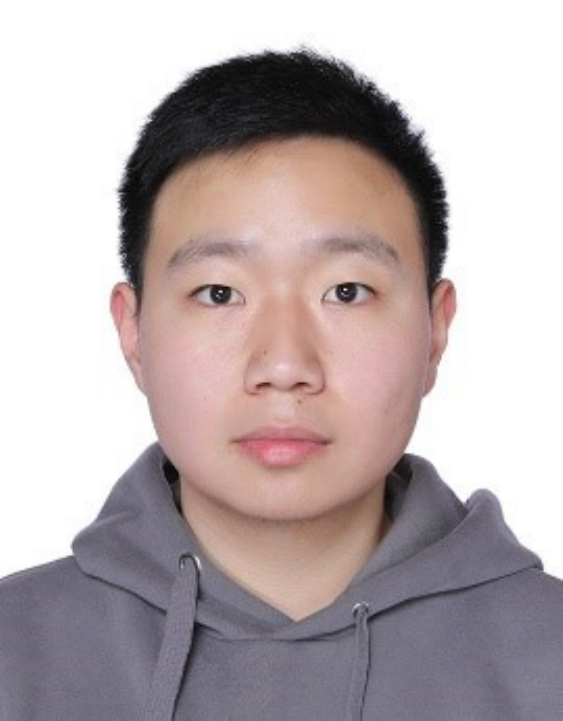}}]{Ruichen Zhang} received the B.E. degree from Henan University (HENU), Kaifeng, China, in 2018. He received his Ph.D. degree from Beijing Jiaotong University (BJTU), Beijing, China, in 2023. He is now working as a Postdoctoral Research Fellow with the College of Computing and Data Science at Nanyang Technological University (NTU), Singapore. His research interests include reinforcement learning-enabled wireless communication networks, generative AI-enabled wireless communication networks, and heterogeneous wireless networks. He serves as a reviewer for IEEE Journal on Selected Areas in Communications, IEEE Transactions on Mobile Computing, IEEE Transactions on Wireless Communications, IEEE Transactions on Vehicular Technology, IEEE Internet of Things Journal, IEEE Wireless Communication Letters, etc.
\end{IEEEbiography}

\begin{IEEEbiography}[{\includegraphics[width=1in,height=1.25in,clip,keepaspectratio]{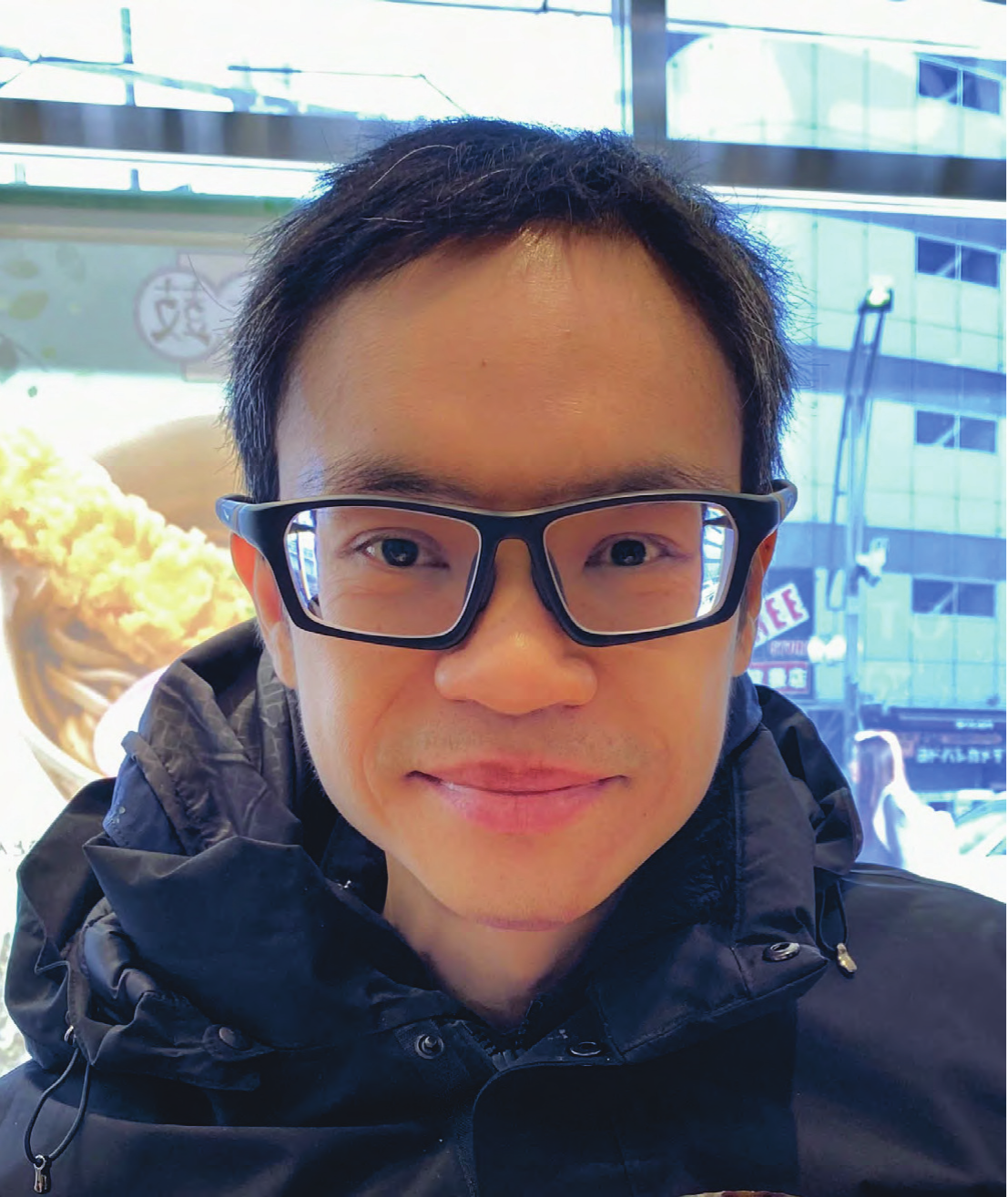}}]{Dusit Niyato} (Fellow, IEEE) is currently a professor in the School of Computer Science and Engineering, at Nanyang Technological University, Singapore. He received B.Eng. from King Mongkut's Institute of Technology Ladkrabang (KMITL), Thailand in 1999 and the Ph.D. in Electrical and Computer Engineering from the University of Manitoba, Canada in 2008. His research interests are in the areas of the Internet of Things (IoT), machine learning, and incentive mechanism design.
\end{IEEEbiography}

\begin{IEEEbiography}[{\includegraphics[width=1in,height=1.25in,clip,keepaspectratio]{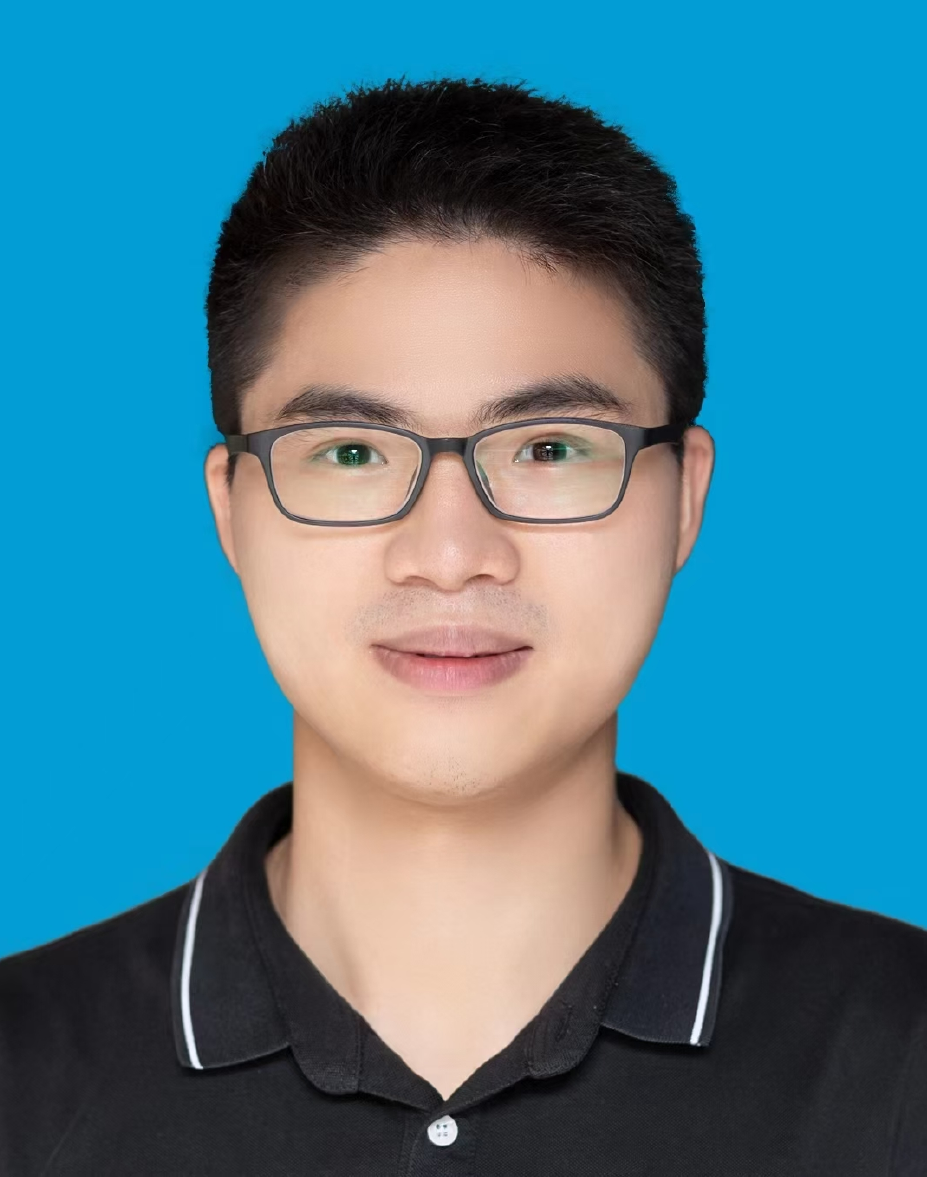}}]{Jiawen Kang}
	received the Ph.D. degree from the Guangdong University of Technology, China in 2018. He was a postdoc at Nanyang Technological University, Singapore from 2018 to 2021. He currently is a full professor at Guangdong University of Technology, China. His research interests mainly focus on blockchain, security, and privacy protection in wireless communications and networking.
\end{IEEEbiography}

\begin{IEEEbiography}[{\includegraphics[width=1in,height=1.25in,clip,keepaspectratio]{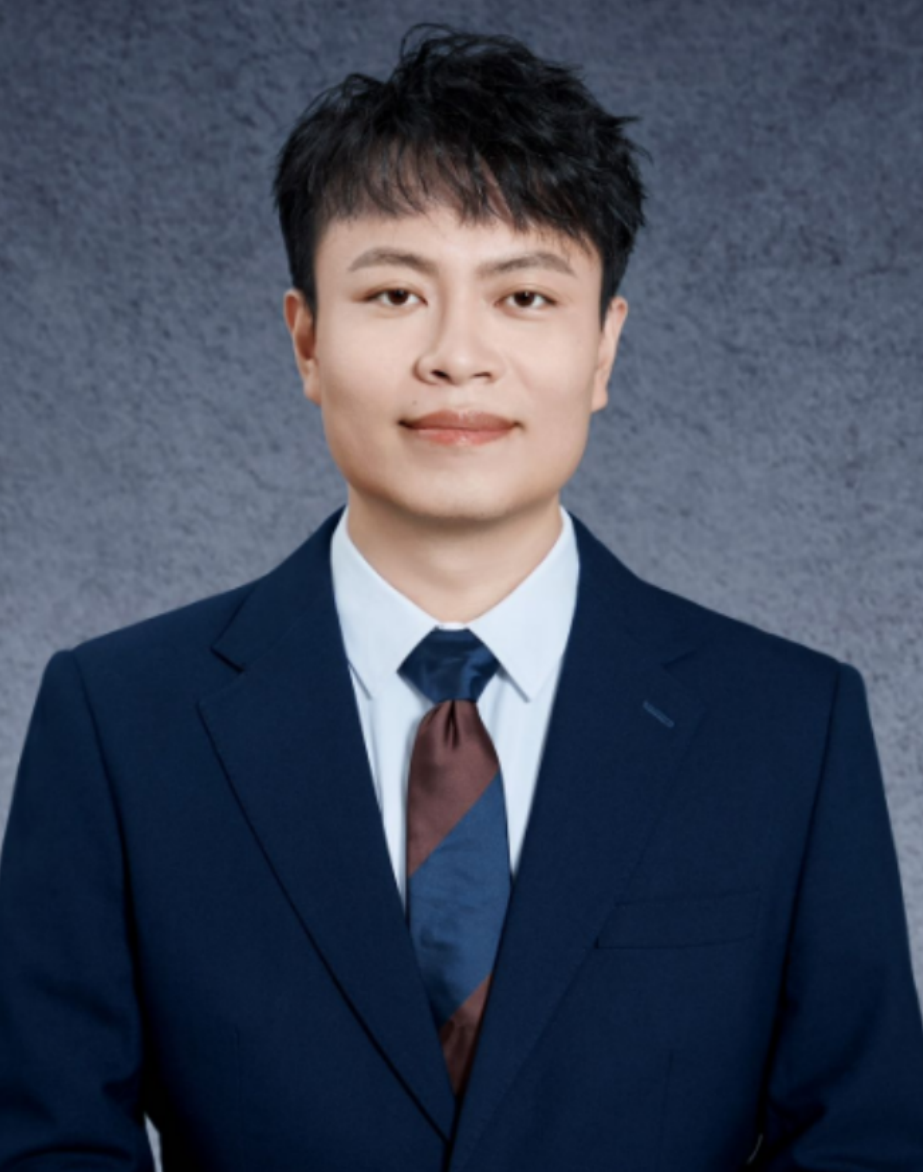}}]{Zehui~Xiong} is an Assistant Professor at Singapore University of Technology and Design (SUTD), and also an Honorary Adjunct Senior Research Scientist with Alibaba-NTU Singapore Joint Research Institute, Singapore. He obtained the B.Eng degree with the highest honors in Telecommunications Engineering at Huazhong University of Science and Technology (HUST), Wuhan, China. He received the Ph.D. degree in Computer Science and Engineering at Nanyang Technological University (NTU), Singapore. He was a visiting scholar with Department of Electrical Engineering at Princeton University and a visiting scholar with Broadband Communications Research (BBCR) Lab in Department of Electrical and Computer Engineering at University of Waterloo. His research interests include wireless communications, Internet of Things, blockchain, edge intelligence, and Metaverse.  Recognized as a Highly Cited Researcher, he has published more than 200 peer-reviewed research papers in leading journals and flagship conferences. He has won over 10 Best Paper Awards in international conferences. In 2023, he was featured on the list of Forbes Asia 30 under 30. He is now serving as the Associate Director of Future Communications R\&D Programme.
\end{IEEEbiography}

\begin{IEEEbiography}[{\includegraphics[width=1in,height=1.25in,clip,keepaspectratio]{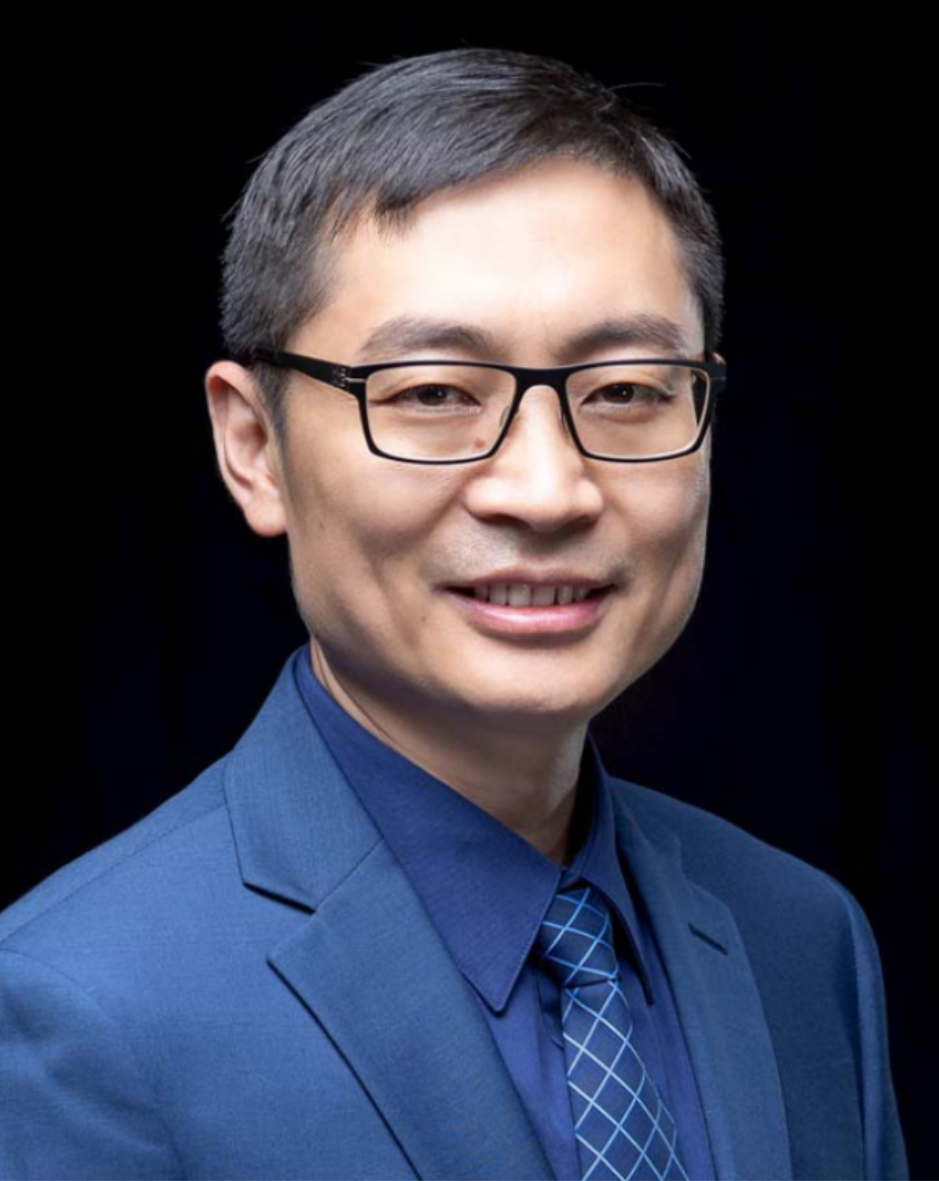}}]{Shuguang Cui} (Fellow, IEEE) received the Ph.D. degree in electrical engineering from Stanford University, CA, USA, in 2005. Afterwards, he has been working as an Assistant Professor, an Associate Professor, a Full Professor, and a Chair Professor in electrical and computer engineering with the University of Arizona, Texas A\&M University, UC Davis, and CUHK at Shenzhen, respectively. He has served as the Executive Dean of the School of Science and Engineering, CUHK at Shenzhen, the Executive Vice Director of the Shenzhen Research Institute of Big Data, and the Director of the Future Network of Intelligence Institute (FNii). His current research interests focus on the merging between AI and communication networks. He was selected as the Thomson Reuters Highly Cited Researcher and listed in the Worlds' Most Influential Scientific Minds by ScienceWatch in 2014. He was a recipient of the IEEE Signal Processing Society 2012 Best Paper Award. He has served as the general co-chair and the TPC co-chairs for many IEEE conferences. He has been an elected member of IEEE Signal Processing Society SPCOM Technical Committee (2009–2014) and the elected Chair of IEEE ComSoc Wireless Technical Committee (2017–2018). He is a member of the Steering Committee of IEEE TRANSACTIONS ON BIG DATA and the Chair of the Steering Committee of IEEE TRANSACTIONS ON COGNITIVE COMMUNICATIONS AND NETWORKING. He has also been serving as an Area Editor for IEEE Signal Processing Magazine and an Associate Editor for IEEE TRANSACTIONS ON BIG DATA, IEEE TRANSACTIONS ON SIGNAL PROCESSING, IEEE JOURNAL ON SELECTED AREAS IN COMMUNICATIONS Series on Green Communications and Networking, and IEEE TRANSACTIONS ON WIRELESS COMMUNICATIONS. In 2020, he won the IEEE ICC Best Paper Award, ICIP Best Paper Finalist, and the IEEE Globecom Best Paper Award. In 2021, he won the IEEE WCNC Best Paper Award. In 2023, he won the IEEE Marconi Best Paper Award, got elected as a fellow of Canadian Academy of Engineering, and starts to serve as the Editor-in-Chief for IEEE TRANSACTIONS ON MOBILE COMPUTING.
\end{IEEEbiography}


\begin{IEEEbiography}[{\includegraphics[width=1in,height=1.2in,clip,keepaspectratio]{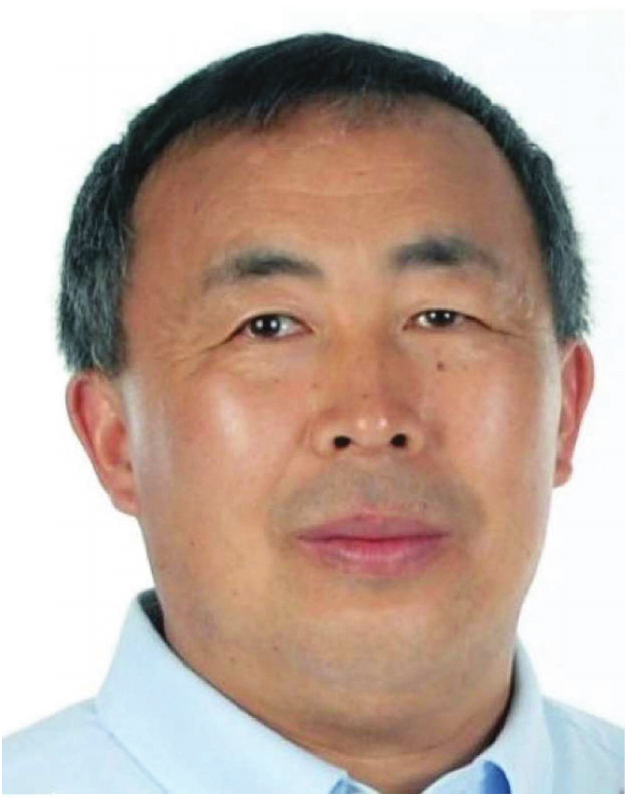}}]{Xuemin (Sherman) Shen} received the Ph.D. degree in electrical engineering from Rutgers University, New Brunswick, NJ, USA, in 1990. He is a University Professor with the Department of Electrical and Computer Engineering, University of Waterloo, Canada. His research focuses on network resource management, wireless network security, Internet of Things, AI for networks, and vehicular networks. Dr. Shen is a registered Professional Engineer of Ontario, Canada, an Engineering Institute of Canada Fellow, a Canadian Academy of Engineering Fellow, a Royal Society of Canada Fellow, a Chinese Academy of Engineering Foreign Member, an International Fellow of the Engineering Academy of Japan, and a Distinguished Lecturer of the IEEE Vehicular Technology Society and Communications Society. Dr. Shen received ``West Lake Friendship Award'' from Zhejiang Province in 2023, President's Excellence in Research from the University of Waterloo in 2022, the Canadian Award for Telecommunications Research from the Canadian Society of Information Theory (CSIT) in 2021, the R.A. Fessenden Award in 2019 from IEEE, Canada, Award of Merit from the Federation of Chinese Canadian Professionals (Ontario) in 2019, James Evans Avant Garde Award in 2018 from the IEEE Vehicular Technology Society, Joseph LoCicero Award in 2015 and Education Award in 2017 from the IEEE Communications Society (ComSoc), and Technical Recognition Award from Wireless Communications Technical Committee (2019) and AHSN Technical Committee (2013). He has also received the Excellent Graduate Supervision Award in 2006 from the University of Waterloo and the Premier’s Research Excellence Award (PREA) in 2003 from the Province of Ontario, Canada. Dr. Shen is the Past President of the IEEE Communications Society. He was the Vice President for Technical \& Educational Activities, Vice President for Publications, Member-at-Large on the Board of Governors, Chair of the Distinguished Lecturer Selection Committee, and Member of IEEE Fellow Selection Committee of the ComSoc. Dr. Shen served as the Editor-in-Chief of the IEEE IoT JOURNAL, IEEE Network, and IET Communications.
\end{IEEEbiography}

\begin{IEEEbiography}[{\includegraphics[width=1in,height=1.25in,clip,keepaspectratio]{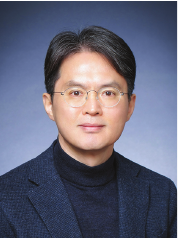}}]{Dong In Kim} (Fellow, IEEE) received the Ph.D. degree in electrical engineering from the University of Southern California, Los Angeles, CA, USA, in 1990. He was a Tenured Professor with the School of Engineering Science, Simon Fraser University, Burnaby, BC, Canada. He is currently a Distinguished Professor with the College of Information and Communication Engineering, Sungkyunkwan University, Suwon, South Korea. He is a fellow of the Korean Academy of Science and Technology and a member of the National Academy of Engineering of Korea. He was the first recipient of the NRF of Korea Engineering Research Center in Wireless Communications for RF Energy Harvesting from 2014 to 2021. He received several research awards, including the 2023 IEEE ComSoc Best Survey Paper Award and the 2022 IEEE Best Land Transportation Paper Award. He was selected the 2019 recipient of the IEEE ComSoc Joseph LoCicero Award for Exemplary Service to Publications. He was the General Chair of the IEEE ICC 2022, Seoul. Since 2001, he has been serving as an Editor, an Editor at Large, and an Area Editor of Wireless Communications I for IEEE Transactions on Communications. From 2002 to 2011, he served as an Editor and a Founding Area Editor of Cross-Layer Design and Optimization for IEEE Transactions on Wireless Communications. From 2008 to 2011, he served as the Co-Editor-in-Chief for the IEEE/KICS Journal of Communications and Networks. He served as the Founding Editor-in-Chief for the IEEE Wireless Communications Letters from 2012 to 2015. He has been listed as a 2020/2022 Highly Cited Researcher by Clarivate Analytics.
\end{IEEEbiography}

\end{document}